\documentclass{ar}
\jname{Annu. Rev. Nucl. Part. 2017.67:129-59}
\jvol{AA}
\jyear{2017}
\doi{10.1146/annurev-nucl-102115-044939}

\def\gtorder{\mathrel{\raise.3ex\hbox{$>$}\mkern-14mu
 \lower0.7ex\hbox{$\sim$}}}
\def\ltorder{\mathrel{\raise.3ex\hbox{$<$}\mkern-14mu
 \lower0.7ex\hbox{$\sim$}}}

\usepackage{caption}

\begin{document}
\markboth{Fomin et al.}{New Results}
\title{New Results on Short-Range Correlations in Nuclei }

\author{Nadia Fomin$^1$, Douglas Higinbotham$^2$, Misak Sargsian$^3$ 
and Patricia Solvignon$^{4*}$
\affil{$^1$ University of Tennessee, Knoxville, TN 37996}
\affil{$^2$ Jefferson Lab, Newport News, VA, 23606}
\affil{$^3$ Florida International University, Miami, FL 33199}
\affil{$^4$ University of New Hampshire, Durham, NH 03824}}

 \begin{abstract}
Nuclear dynamics at short distances is one of the most fascinating topics of strong interaction physics. 
The physics of it is closely related to the understanding the role of the QCD in generating  nuclear forces 
at short distances as well as understanding the dynamics of the super-dense cold nuclear matter relevant to 
the interior of neutron stars.  With an emergence of high energy  electron  and proton beams there is a significant recent progress
in high energy nuclear scattering experiments aimed at studies of short-range structure of nuclei.  This in turn stimulated 
new theoretical studies resulting in the observation of several new phenomena specific to the short range structure of nuclei.
In this work we review recent theoretical and experimental progress in studies of short-range correlations in nuclei and their 
importance for advancing  our  understanding of the dynamics of nuclear interactions at small distances. 
% Abstract text, approximately 150 words. 
\end{abstract}
\begin{keywords}
Short-range nucleon correlations, high energy electron nucleus scattering, deep inelastic nuclear scattering,
super-fast quarks
%keywords, separated by comma, no full stop, lowercase
\end{keywords}
\maketitle

%\date{\today}     
%\maketitle

%\sections{Introduction}
%\input{1introduction}

\section{Introduction}
%\noindent  \fbox{\bf 1.5 pages}\\

\noindent {\bf Nuclear Physics at Short Distances:} One of the outstanding issues in contemporary nuclear  physics 
is understanding the role of Quantum Chromodynamics (QCD) in generating the nuclear forces.  There has been significant 
progress in understanding the medium to long range ($\gtorder 1.2$~fm)  part of the nuclear forces in terms  of effective 
field theories in which pion exchange forces are constructed to satisfy QCD symmetries and the dynamics in which pion is a Goldstone 
boson of the spontaneously broken chiral symmetry (for recent review see Ref.~\cite{Machleidt:2016rvv}).  
However,  one of the challenging and less understood domains of nuclear forces is that of intermediate to short 
distances ($\ltorder 1.2$~fm).   This is the domain of the transition from baryonic to 
quark-gluon degrees of freedom where one expects a host of  new phenomena inherent to QCD dynamics  which 
are not visible to long range nuclear forces.  These include chiral symmetry, quark interchanges,  gluon dynamics, 
hidden color states, among others.  One of the most fascinating short distance phenomena  is the practically unexplored 
dynamics of the nuclear repulsive core.  We owe our existence literally  to the nuclear core, without which nuclei 
will collapse to  sizes $\le 0.3-0.4$~fm,   followed by  the onset of quark-gluon degrees of freedom and 
restoration of chiral symmetry with rather unimaginable  consequences for the order of the universe we know.

Within above mentioned effective field theories the short-range nuclear forces are described through the 
effective contact interactions whose internal structure is invisible in low energy processes. However, in high momentum transfer processes
the dynamical content of these contact interactions are increasingly important and the current  review addresses the recent progress 
in these studies.
From the nuclear physics point of view this is the region dominated by short range multi-nucleon correlations. 
The understanding of the  dynamics of these correlations has important relevance not only to the 
QCD dynamics of nuclear forces  but also to the physics of high density nuclear matter that  one expects to exist at the 
cores of neutron stars.

\medskip

\noindent{\bf Importance of Short-Range Correlations:}
The atomic nucleus, as a bound system of interacting fermions with the Hamiltonian not 
commuting with the momentum operator, creates the phenomenon of momentum distribution of its constituents in the nucleus. 
Fermi statistics, combined with the fact that the combined volume of  A nucleons is comparable to the nuclear volume, makes 
the effect of the nucleon correlations a phenomenon that ranges from the long to short distances.   
Long range correlations are reasonably well studied experimentally and have been  understood based on Pauli  
blocking effects and the long-range part of the nuclear forces~\cite{Lapikas:1999ss}.
Short-range nucleon correlations, one the other hand, represent the most intriguing part of the nuclear dynamics whose experimental 
studies are intensifying with the emergence of high energy accelerators capable of performing nuclear experiments with high intensity 
beams, allowing for precision measurements of small cross-sections.

Historically, the importance of short range correlations~(SRCs) was first emphasized by R.~Jastrow~\cite{Jastrow:1955zz} who observed 
that application of variational methods to estimate the upper limit of the ground state energy fails for the problem involving strong 
interaction whose range, $r_0$,  is comparable with the size of the nucleons.  In this case, the expectation value of the mean potential 
energy per nucleon is: $\bar V = {Nr_0^3\over \Omega} \mid V_0\mid$,  with $\Omega$ being the nuclear volume and $V_0$ the average potential 
of the $NN$ interaction at $r\le r_0$. From this relation one observes that in the hard sphere limit of the NN interaction core, the upper 
bound of the ground state energy increases without limit.   This problem was solved~\cite{Jastrow:1955zz} by introducing  NN correlations 
at $r\le r_0$ distances based on the cluster decomposition approach choosing an ad-hoc form for the correlation function.   While this approach 
solved the ground state energy problem it did not address the dynamical origin of such correlations. 
About the same time the correlation functions were introduced within the framework of the Brueckner - Goldstone theory 
(for the  review and  relevant references see Ref.\cite{Bethe:1971xm}), in which the nucleon's finite size  introduced 
a positional correlations that modified the independent particle state wave function.

An approach which demonstrated how the dynamical property of strong forces at short distances enters into SRCs was illustrated 
in Ref.~\cite{Amado:1976zz} by obtaining the asymptotic solution of the N-body Shroedinger equation in  the large momentum  limit 
of the bound nucleons. In this case  for specific behavior of the  NN potential at short distances (see below) the momentum distribution 
of high momentum nucleon is defined as $n(p)\sim {\mid V(p)\mid^2\over p^4}$.  This relation indicates  that probing large momentum of the 
bound  nucleon in the nucleus allows to access the NN potential at large relative momenta or small relative distances.
It is worth mentioning that the observation of the relation between  short distance phenomena in the nucleus and 
high momentum component of nuclear wave function was already made within Brueckner-Goldston theory (see e.g. Ref.\cite{Brueckner:1955zzd}) however 
the analytic property of NN potential was not identified for which the high momentum  component would have the above presented 
form.

This approach defines the {\em main strategy} of SRC studies,  which is, accessing the dynamics of  NN interactions 
at short distances by probing bound nucleons with large initial momentum and removal energy.
 
Another issue that SRC studies aim to address is the existence of 3N and possibly higher order  correlations.
The questions in this case are whether SRCs of more than two nucleons can be formed and the necessary conditions 
to generate deeply bound high-momentum nucleons.
These studies have significant astrophysical importance in understanding the dynamics of the super-dense 
nuclear  matter which is formed in the core of the neutron stars.

Finally, a new direction of  SRC studies is to probe the QCD content of these high density fluctuations.  
Recent experiments have begun to probe quark distributions in the kinematic region in which single quark 
carries momentum larger than the individual nucleon and are referred to as super-fast quarks. 
 
 \medskip

\noindent {\bf  Experiments  with High Energy  and  High Intensity Beams:}As introduced above, the main experimental methodology 
of accessing SRCs is to probe a deeply bound nucleon in the nucleus with large momentum and removal energy.  To realize such a 
possibility one needs  processes in which the energy   transferred to the nucleon in  the SRC  significantly exceeds the  potential 
and kinetic energies characteristic of the SRC.  Such an instantaneous removal of the deeply bound nucleon from the SRC will  
release  correlated  nucleons, the detection of which will provide an additiona  window into the dynamics of SRCs in the nucleus.

Such experiments are only possible with high energy probes.
The first possibility presented itself at  Stanford Linear Accelerator Center~(SLAC) where electron beams with 
up to $20$~GeV energy  made it possible to perform unprecedented measurements of inclusive $A(e,e')X$ processes 
in the kinematic region dominated by SRCs~\cite{Rock:1982gf,Arnold:1988us,Rock:1991jy,Day:1979bx,Rock:1981aa}.  
These experiments allowed, for the first time, to reach the high-momentum component of nuclear wave function through 
the $y$-scaling analysis\cite{Day:1987az}.  Subsequently, the study\~cite{frankfurt93} of the ratios of inclusive cross 
sections of nuclei $A$ to the deuteron revealed the first signatures of two-nucleon SRCs in the form of the scaling as a 
function of  Bjorken variable $x$ in the region of $1<x<2$.

Meanwhile, attempts  were made to extend the SRC studies to semi-inclusive processes in which the knock-out proton or 
the proton emerging from the 2N SRCs as a spectator,  were detected in the coincidence with the scattered 
electron~\cite{Royer:1975zz,Marchand:1987hd,Alanakian:1997pm,Alanakian:1998qr}.  These experiments observed 
unambiguously the SRC signatures in the form of correlations between missing energy and 
missing momentum~\cite{Marchand:1987hd} or through the shifts of the scattered electron spectra proportional to  
the energy and angle of spectator protons~\cite{Alanakian:1997pm,Alanakian:1998qr}.  
However, such signatures are present for any two-body  currents and due to restricted magnitudes 
of transferred energy and momentum it was impossible to dynamically  suppress the soft two-body processes  
not related to SRCs (such as meson exchange currents).

The situation was completely changed with the emergence of CEBAF's 6~GeV continuous electron beam at Jefferson Lab (JLab), 
where practically all the  major advances in SRC studies have been made in recent years.  JLab's high energies along with high
intensities of the electron beam allowed precision measurements of small cross-sections characteristic of SRC dynamics in electroproduction reactions. 
Additionally, triple coincidence experiments were carried out where additional nucleons were detected: one struck from the SRC by a virtual 
photon and the other emerged as a spectator from the same SRC.
It is worth mentioning that about the time when JLab experiments were producing first results,SRC 
studies were also performed at Brookhaven National Laboratory~(BNL). In the BNL experiments the AGS was used as a source 
for the  high energy proton beams (6-15~GeV/c) to perform double $A(p,2p)X$ and triple $A(p,2pn)X$ coincident experiments
aimed at SRC studies.  Measurements of both electro-nuclear and proton-nuclear experiments were remarkable as  
they yielded  almost identical results on the dynamics of SRCs, thereby confirming the universal nature of the object they were probing in the nucleus. 

These experiments further stimulated the theoretical research on SRCs, covering issues related to high energy electro-nucleon 
processes, modeling high-momentum components of nuclear wave function, understanding the dominance of proton-neutron component 
in the SRCs and understanding of the role of the SRCs in medium modification of the bound nucleons.   These together with the recent 
results in experimental studies of SRCs  are the subject of the current review.
 
\medskip

\noindent{\bf Summary of the Recent Progress:}
The series of recent experiments studying  high energy $eA$ and $pA$ 
processes~\cite{egiyan03,egiyan06,fomin2012,piasetzky06,shneor07,subedi08,Hen:2014nza} 
have greatly improved our understanding of the dynamics of 2N SRCs in nuclei. 
The new generation of inclusive  $A(e,e^\prime)X$ experiments\cite{egiyan03,egiyan06,fomin2012} confirmed the 
observation made from  the analysis of SLAC data\cite{frankfurt93,frankfurt88} that the ratios of nuclear 
to deuteron cross sections scale  in the kinematic region dominated by  the scattering from the 2N SRC. 
The ratios, designated by the parameter, $a_2(A,Z)$, directly related to the SRC strength in the high-momentum 
component of nuclear wave functions were measured for a wide range of $A$.

Analysis of the $A(e,e^\prime)X$ data in the deep-inelastic region at Jefferson Lab yielded EMC ratios which 
characterize the extent of  medium modification of partonic  distributions of the bound nucleon~\cite{Seely:2009gt}.
Comparison of the strengths  of the EMC ratios with the $a_2(A,Z)$ parameters  revealed an
apparent correlation between these two observables~\cite{Weinstein:2010rt,Arrington:2012ax},  strongly suggesting that the medium modification effect is related to the probability of nucleons being in short-range correlations.

The final group of SRC experiments consisted of high-energy, semi-inclusive triple-coincidence  measurements~\cite{piasetzky06,subedi08}, 
which succeeded in probing the  isospin composition of  2N SRCs in the relative  momentum range  of $\sim 250-650$~MeV/c. These experiments 
observed a  strong  (by factor of 20) dominance  of the  proton-neutron SRCs in nuclei as compared to  the proton-proton  and neutron-neutron  correlations.  
Such an excess is understood~\cite{piasetzky06,sargsian05,schiavilla06} based on the dominance of the tensor forces in the NN interaction 
at the above mentioned momentum range corresponding to the average inter-nucleon separations of $\sim 1.1$~fm.   

Based on the observation of the strong dominance of $pn$ SRCs, it  was  predicted in Ref.~\cite{Sargsian:2012sm,McGauley:2011qc} 
that single proton or neutron  momentum distributions in the 2N SRC domain are inversely proportional to their relative fractions 
in  nuclei.  This prediction is in agreement with the results of  variational Monte-Carlo calculation of momentum distributions of  
light nuclei~\cite{Wiringa:2013ala} as well with the SRC model calculation for medium to heavy nuclei~\cite{Vanhalst:2014cqa}.  
The recent experimental verification of the $pn$ dominance in heavy nuclei (up to $^{208}Pb$)~\cite{Hen:2014nza} provides  
strong evidence for the universality of the above prediction across nuclei.

The isosinglet $pn$ dominance in the SRC region makes the dedicated   studies of the high-momentum component of the deuteron 
wave function a priority. The first experiment aimed at probing deuteron at large internal momenta was completed recently at 
JLab~\cite{Boeglin:2011mt}.  Here, the exclusive $d(e,e'p)n$ cross section was measured at unprecedentedly large  $Q^2=3.5$~GeV$^2$. 
This experiment succeeded for the first time in measuring deuteron momentum distribution for up to $550$~MeV/c  with negligible and controlled 
effects due to long-range nuclear phenomena and final state interactions.  This experiment will be extended to even more extreme
kinematics with the upgraded 12~GeV JLab accelerator.
The experimental progress summarized above has stimulated strong  theoretical activity in understanding the dynamical origin of 
short-range correlations and their implications in the different nuclear phenomena, including the EMC effect, 
symmetry energy and the nuclear 
contact~\cite{Hen:2016kwk,Alvioli:2016wwp,Chen:2016bde,Boeglin:2015cha,Neff:2015xda,Atti:2015eda,Cai:2015xga,Furnstahl:2013oba,Weiss:2016obx,CiofidegliAtti:2017tnm}.

%\section{Recent Advances}
%\input{2recent_advances}

\section{Recent Advances in Theory}

\subsection{Emergence of Short Range Correlations in Strongly Interacting Fermi Systems}
\label{emergenceSRC}

While in Jastrow's paper~\cite{Jastrow:1955zz} the correlation picture of multi-nucleon system was discussed within 
cluster development approach, it did not address the dynamical origin of the correlations.   
Such a discussion first was made in Ref.\cite{Amado:1976zz} where it was shown that in the limit of very large 
internal momenta of the bound nucleon, the solution of the Lipman-Schwinger  equation is defined at  large relative momentum 
of the pair-wise $NN$ interaction potential. However, this work did not discuss whether 
the two-nucleon  correlations will dominate the large momentum component of the nuclear wave function relative to 
the higher order (more than two nucleon) correlations.  Subsequent works~\cite{frankfurt81}  analyzed the issue of 
hierarchy of NN correlations and  observed 
that if the NN potential is a finite-range potential  and in the high-momentum ($q\gg m_N$)
limit  behaves as $V(q) \sim {1\over q^n}$ (with $n>1$), then 
the wave function of the multinucleon system in the limit where one of the nucleons has a large momentum, is dominated mainly by 
two-nucleon correlations.

One can see such an emergence   of 2N SRCs considering the Lipmann-Schwinger equation  for 
a A-nucleon bound system interacting through the finite-range pair-wise  $NN$ interactions,  $V_{i,j}$,  in the form
\begin{equation}
\phi_A(k_1,\cdots k_n, \cdots k_A) =  { -{1\over 2}\sum\limits_{i\ne j} V_{ij}(q) \phi_A(k_1,\cdots, k_i+q,\cdots, k_j-q,\cdots, k_A){d^3q\over (2\pi)^3}\over
 \sum\limits_{i=1}^{A}{k_i^2\over 2m_N} - E_B},
 \label{LSch}
\end{equation}
where this equation can be used in the iteration approach to calculate the full nuclear wave function.
If one starts with the first iteration using the mean-field nuclear wave function in the RHS part of the equation 
and  considers  one of the  bound nucleons (say $j$) in the LHS part   having momentum $p$ such that ${p^2\over 2m_N}\gg E_B$,
one observes that the integral in the RHS part  will be dominated by the configuration in which 
$\bf q\approx \bf p$ such that $\bf {k_j} - {\bf q}  \approx 0$ and ${\bf k_i} \approx - {\bf k_j} \approx  -{\bf p}$ such that \textbf{$k_i+q\approx 0$}. 
This situation  results in  a two-nucleon correlation with large relative momentum such that~\cite{Amado:1976zz}:
\begin{equation}
\phi^{(1)}_A(k_1,\cdots,k_i= p,\cdots, k_j \approx -p, \cdots,k_A) \sim {V_{NN}(p)\over p^2} f(k_1,\cdots^\prime \cdots ^\prime\cdots  k_A),
\label{it1}
\end{equation}
where $f(\cdots)$ is a smooth function of the momenta of non-correlated nucleons not containing momenta $k_i$ and $k_j$.
However this result does not yet guarantee the dominance of  $2N$ correlations in  the high momentum part of the nuclear wave function.
In order for this to happen, higher order correlations should be parametrically small. In the approximation of pair-wise interaction the higher order correlations correspond to the higher order iterations.  For example, 3N SRCs can be estimated
 by  inserting   Eq.(\ref{it1}) into  Eq.(\ref{LSch}), yielding:
\begin{equation}
\phi^{(2)}_A(\cdots p, \cdots) \sim {1\over p^2}\int {V_{NN}(q) V_{NN}(p)\over (p-q)^2} d^3 q. 
\label{it2}
\end{equation}
If now one assumes analytic behavior  for the $NN$ potential of the form $V_{NN}(q) \sim q^{-n}$ in the large $q$ limit, then one estimates the parametric dependence of the wave function due to three nucleon SRCs as:
\begin{equation}
\phi^{(2)}_A(\cdots p, \cdots) \sim {V(p)\over p^2}\int\limits_{q_{min}}^{\infty} {dq \over q^n},
\end{equation}
where $q_{min}$ is due to finite range of the potential.  This relation indicates that for the finite-range interaction with $n>1$, 
the second (as well as higher order) correlations will be parametrically  suppressed (see also Refs.~\cite{frankfurt88,piasetzky06}).
 
An important feature  of the above  result is that one will also arrive at the same conclusion in the relativistic framework if the nuclear 
dynamics are described within light-front or  Weinberg-type equations for the bound systems (see e.g. \cite{Weinberg:1966jm,frankfurt81}).
 
The above discussion allows us to conclude that bound systems interacting through the pair-wise  Yukawa type interactions 
will generate correlations dominated by  two nucleon $NN$ SRCs.  Higher order correlations are possible but they are parametrically 
suppressed relative to 2N SRCs.

To access such correlations experimentally, one needs to probe bound nucleons with high initial momenta, requiring ${p^2\over 2m_N}\gg  E_B$. If such conditions are met, then the  asymptotic  form of Eq.(\ref{it1})  leads to the following approximate relation for a nucleon momentum distribution  
at $p> k_{F}$, with $k_F$ being the characteristic  Fermi momentum of the nucleus: 
 \begin{equation}
 n^A({p}) \sim a_{NN}(A) \cdot n_{NN}({p}),
\label{2NSRCmodel}
 \end{equation}
 where the full momentum distribution is normalized as $\int n^A({p})d^3p = 1$. The parameter  $a_{NN}(A)$ can be interpreted as a probability 
 of finding NN SRC in the given nucleus $A$. The function, $n_{NN}({p})$ is the momentum distribution 
 in the NN SRC.

\subsection{Conceptual Issues in High Energy Nuclear Theory}
\label{sec2.2}

\subsubsection{Emergence of Light-Front Dynamics}
The above discussion indicates that probing  short distance correlation structure of the nucleus requires probing 
a bound nucleon with momentum significantly exceeding the characteristic Fermi  momentum, $k_{F} \sim 250~MeV/c$. 
To achieve an instantaneous removal of such a nucleon from the nucleus, the momentum, $q$,  transferred to the bound nucleon
should be sufficiently large: $q \gg  2k_{F}$.  Another restriction to the momentum transfer follows from the dynamic picture of SRCs, in which 
the correlated nucleons have large and comparable momenta in opposite directions.  Therefore,  in order to unambiguously discriminate the struck nucleon in the final state with momentum ${\bf p_f} = {\bf p_i} + {\bf q}$   from the correlated spectator nucleon ~($p_s$),  one needs to satisfy the condition of:
\begin{equation}
{q} \sim { p_f} \gg p_s \sim  300 - 1000 \ \mbox{MeV/c}
\label{qcon}
\end{equation}
Requiring that the interaction off the bound nucleon is quasi-elastic allows us to determine 
${\bf p_i}$ and leads to the kinematic threshold of $Q^2 = {\bf q}^2 - q_0^2 > m_N^2$. 
Thus one arrives at the optimal kinematics for probing SRC structure of nuclei as:
\begin{equation}
Q^2 > m_N^2, \ \ \ q \sim p_f \ge \mbox {few GeV/c},  \ \ \ p_s \sim \mbox {few hundred MeV/c}
\label{highkin}
\end{equation}
These  kinematic conditions  bring us to the domain of high energy physics, meaning that the constituent masses of the 
bound system become increasingly less important.  From the theoretical point of view, such a condition indicates an increased validity of the methods 
of high energy physics~\cite{Feynman:1973xc} associated with the onset of the light-front dynamics and calculational methods 
utilizing a new kind of (kinematic) small parameter  such as 
\begin{equation}
{q_{-}\over q_{+}}\sim {p_{f-}\over p_{f+}} \ll 1,
\label{smallparameter}
\end{equation} 
where $q_\pm = q_0 \pm {\bf q}$ and $p_{f\pm} = E_f\pm p^z_{f}$, with $z$-axis defined to be the direction of $\bf q$.    
Such a small parameter allows for systematic accounting of the strong off-shell effects in the 
reaction mechanism of the  external probe - bound nucleon scattering~\cite{Sargsian:2001ax} as well as the possibility for 
self-consistent calculation of the final state interaction of the fast struck nucleon with the correlated spectators within
generalized eikonal approximation~\cite{Frankfurt:1996xx,Sargsian:2001ax,Sargsian:2009hf}.

The emergence of light-front dynamics in which the scattering process evolves along the light-cone, $t\sim z$, is similar to that of deep-inelastic scattering~(DIS) from a nucleon in which the parton distribution of nucleons is probed. 
Light-front dynamics  in DIS is the most  natural approach, since due to the light masses of
quarks  one deals with a large contribution from  vacuum fluctuations which can be suppressed in the light-front  or 
infinite  momentum reference frames (see e.g. Refs.~\cite{Feynman:1973xc,Lepage:1980fj}).

Vacuum fluctuations arise also in relation to the high momentum component of the nuclear 
wave function.   With the momentum of the nucleon, $p_i$,  becoming comparable with nucleon masses ($p_i\sim m_N$) the 
vacuum diagrams representing  $N\bar N$ fluctuations become  as important as diagrams representing the "valence" component of nuclear wave function 
(see e.g. Refs.~\cite{frankfurt81,Miller:2000kv}). This is an important problem which should be addressed in 
any theoretical studies aimed at the exploration of the high-momentum component of the  nuclear wave function.

\medskip

Another implication of light-front dynamics is the emergence of the light-cone variables   ($\alpha_i, p_T$) describing different 
nuclear "observables"\footnote{They are not direct observables, but ones that can be extracted 
from different scattering processes involving nuclear targets.} such as the nuclear spectral  or light-front density functions.
The variable, $\alpha_i$, is  analogous to the variable $x$
of  QCD and represents the light cone nuclear momentum fraction carried  by the constituent nucleon and defined as 
\begin{equation}
\alpha_i = {p_{i-}\over p_{A-}/A},
\label{alphai}
\end{equation}
where $p_{i-}$ and $p_{A-}$ are "longitudinal" components of the light-cone momenta of the bound nucleon and nucleus
\footnote{Note hat in our definitions  the direction of the $z$ axis is defined by the direction of the momentum transfer $q$, which 
is opposite to that  of frequently defined in QCD analysis of the  nucleon wave function (see e.g. \cite{Lepage:1980fj}).}.

One of the  advantages  of  ($\alpha_i, p_T$) variables is invariance with respect to Lorentz boosts in the ${\bf q}$ direction. 
Such a feature  allows to formulate  nuclear spectral and density functions in a boost invariant form. These functions are nucleonic 
analogues of unintegrated  partonic distribution functions in QCD and can be systematically extracted 
from the analysis of nuclear processes.  However, the extraction is only possible provided the validity of 
factorization between the reaction mechanism and nuclear light-front momentum distributions as well as closure approximation for 
the final state interaction contribution.  All these factors are analogous to DIS processes and can be systematically taken into 
account in the orders of the small parameter presented in Eq.(\ref{smallparameter}) (see e.g. Ref.\cite{Sargsian:2001ax}).
  
If the light-front spectral or density functions are extracted, they will represent a testing ground for the light-front nuclear wave function,
$\psi_A(\alpha_1,p_{T1}, \alpha_2,p_{T2},\cdots \alpha_A, p_{TA})$, through which 
the nuclear spectral and the density functions are constructed as follows:
\begin{eqnarray}
P_A^N(\alpha_i,p_{Ti},\tilde M_N^2) = \sum\limits_{j=1}^A\int \mid \psi_A(\alpha_1,p_{T1}, \alpha_2,p_{T2},\cdots \alpha_A, p_{TA})\mid^2
\delta(\alpha_i - \alpha_j)\delta^2(p_{Ti}-p_{Tj}) \nonumber \\ 
\times \delta(\tilde M_N^2 - (p_A - \sum\limits_{j\ne i = 1}^A p_j)^2 \prod\limits_{j}^A {d\alpha_j\over \alpha} d^2p_{T_j} \ \ \ \ 
\end{eqnarray}
 and
\begin{equation}
 \rho_{A}^N(\alpha,p_T) = \int P_A^N(\alpha_i,p_{Ti},\tilde M_N^2)  {1\over 2} d\tilde M_N^2
 \end{equation}
 normalized to the baryonic number of the nucleus:
 \begin{equation}
 \sum\limits_{N}^A \int \rho_{A}^N(\alpha,p_T) {d\alpha\over \alpha} d^2p_T = A.
 \end{equation}
 
 It is worth  noting  that $\rho_{A}^N$  can be related to $f_{A}(\alpha,p_T)$ which is analogous to the unintegrated 
 partonic distribution function in QCD, via:
 \begin{equation}
 f_{A}(\alpha,p_T) =  {\rho_{A}^N(\alpha,p_T) \over \alpha}
 \end{equation}
These functions, extracted from the high-momentum transfer semi-inclusive and inclusive processes, can provide a testing ground for the SRC properties of the nuclear wave funciton.

%\noindent  \fbox{\bf  1.5 pages }\\

\subsection{Nuclear Dynamics at Sub-Fermi Distances - $pn$ Dominance}
\label{sec_pn_dom}

Emergence of NN correlations in the short-range nuclear dynamics creates a possibility of observing 
a host of new phenomena related to the rich structure of nucleon-nucleon interaction at short distances. One such effect arises from the interplay of  the central and tensor parts of the NN potential, in which due to the repulsive core,
the central part of the potential changes its sign at $\sim 1$~fm  to become repulsive while no 
such transition exists for the tensor part of the potential. As a result,  at  NN separations of $r_{NN}\sim 1\pm 0.2$~fm there is 
a clear dominance of the tensor part of the NN interaction compared to the central potential.

This situation creates an interesting selection rule for the isospin composition of the SRCs.  
The tensor operator does not couple to isotriplet states, i.e. $\hat S_{NN}\mid NN^{I=1}\rangle = 0$, resulting in a situation where the NN SRC is dominated by the isosinglet component of the $pn$ pair.

If the contribution of  $pp$,  $nn$  and isotriplet $pn$  SRCs are negligible, one 
expects that in the momentum region of $\sim k_{F}-600$~MeV/c the momentum distribution in the NN SRC is defined 
by the isosinglet $pn$ correlation only.  Using this fact and the local nature of SRCs one predicts:
\begin{equation}
  n_{NN}({p})\approx n_{pn}({p}) \approx  n_{d}({p}),
 \label{2=d}
 \end{equation}
for Eq.(\ref{2NSRCmodel}), where $n_d({p})$ is the deuteron  momentum distribution at $\sim k_{F}< p \le 600$~MeV/c.

\medskip

\subsubsection{Two New Properties of High Momentum Component of Nuclear Wave Function}

We introduce the individual momentum distributions of 
proton ($n^{A}_{p}({p})$)  and neutron($n^A_n({p})$) such that:
\begin{equation}
n^A({p})  = {Z\over A} n^A_{p}({p}) +   {A-Z\over A}n^A_n({p}),
\label{sum}
\end{equation}
and $\int n^A_{p/n}({p})d^3p = 1$.  Here, the terms in the sum represent the 
probability density of finding a proton or neutron with momentum $p$ in the nucleus.

\medskip

\noindent{\em I. Approximate Scaling Relation:} 
Integrating Eq.(\ref{sum}) within the momentum range of  $\sim k_F-600$~MeV/c
one observes that the terms in the sum give the  total probabilities of finding a proton  and a  neutron in the 
NN SRC.  Since the SRCs, within the approximation where the contributions from the isotriplet NNs are negligible,  consist only of the isosinglet 
$pn$-pairs, the total probabilities  of finding proton and neutron in the SRC are equal.  With the other possibilities for the 
SRC composition neglected, one predicts that in  $\sim k_F-600$~MeV/c region:
\begin{equation}
x_p\cdot n^{A}_{p}({p}) \approx x_n\cdot n^A_n({p}),
\label{p=n}
\end{equation}
where  $x_p = {Z\over A}$, $x_n = {A-Z\over A}$. This represents the {\em first property,
according to which the momentum distributions of proton and neutron weighted by their respective 
fractions are approximately equal.} 

 \begin{figure}[ht]
\centering\includegraphics[width=9cm,height=6cm]{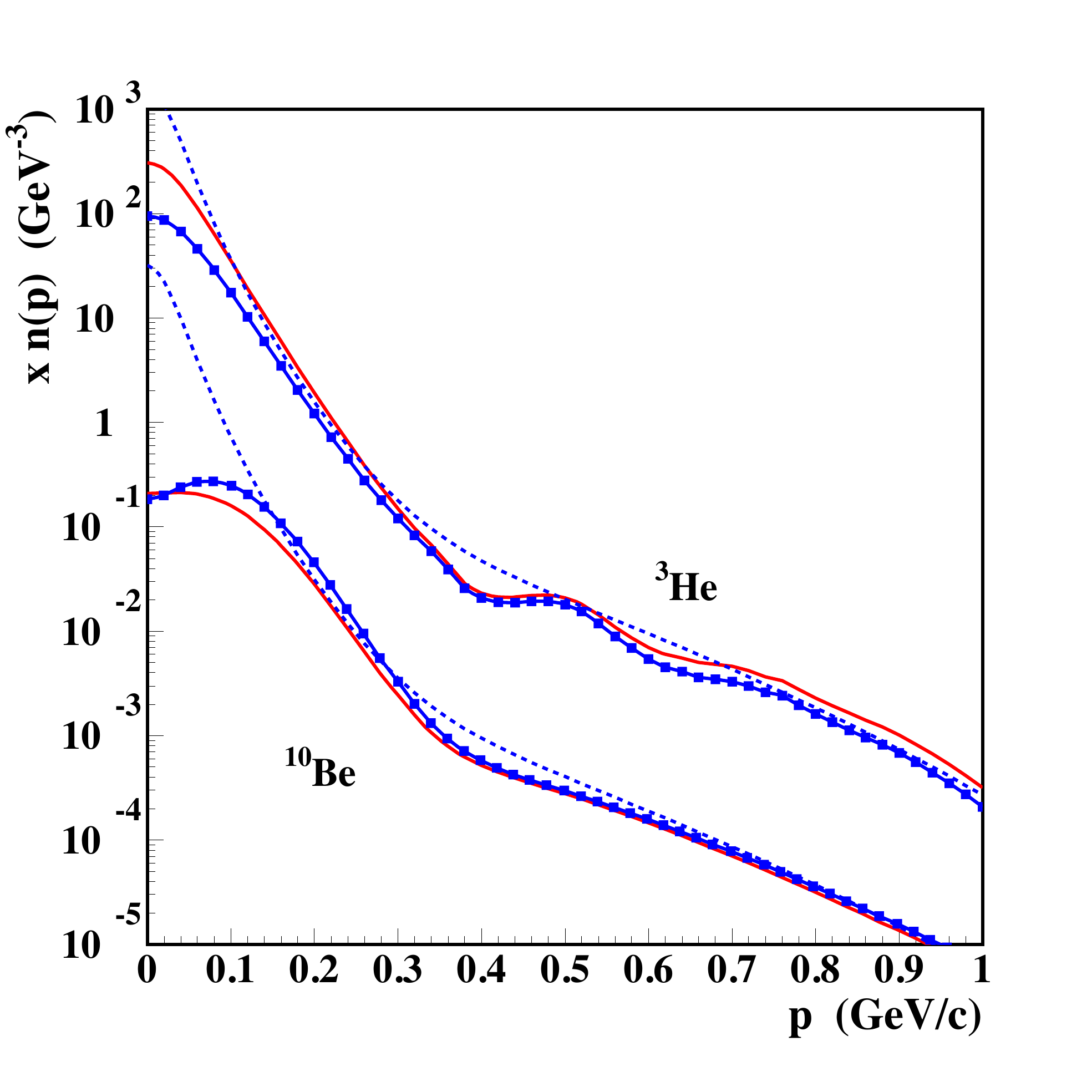}
\caption{(color online) (a) The momentum distributions of proton  and neutron weighted by $x_p$ and $x_n$ respectively.
The doted lines represent the prediction for the  momentum distribution according to Eq.(\ref{highn}).  \ 
 (b)  The $x_{p/n}$ weighted ratio 
of  neutron to  proton  momentum distributions.  See the text for details. Figure  adapted from Ref.\cite{Sargsian:2012sm}.}
\label{He3_Be10}
\end{figure}

The validity of the above approximate scaling rule  is presented in Fig.\ref{He3_Be10} where Eq.(\ref{p=n}) is checked for the  
$^3He$ nucleus using the solution of Faddeev's equation~\cite{Nogga:2000vq},   and 
for $^{10}Be$  using the results of   variational Monte Carlo~(VMC) calculations of Ref.\cite{Wiringa:2013ala}.  
The solid  lines with and without squares in Fig.\ref{He3_Be10}  represent 
neutron and proton momentum distributions for both nuclei weighted by their respective 
${x_n}$ and ${x_p}$ factors.

As can be seen in the figure,  for $^3He$,  the proton momentum distribution 
dominates the neutron  momentum distribution at small momenta  just because there are twice as much protons in $^3He$ and 
no specific selection rules exist for the  mean field momentum distributions.
The same is true for $^{10}Be$ for which now the neutron momentum distribution dominates at small momenta.
However   at  $\sim 300$~MeV/c  for both nuclei, the proton and neutron momentum distributions become close  to each other 
up to the  internal momenta of $600$MeV/c. This is the region dominated by tensor interaction.     
Note that the similar features  present for 
all other asymmetric nuclei calculated  within the VMC method in Ref.\cite{Wiringa:2013ala} for up to $A\le 11$.

\medskip

\noindent{\em II. Fractional Dependence of High Momentum Components:}
Using relations (\ref{2NSRCmodel}) and (\ref{2=d}) for the high momentum distribution $n^A({p})$ 
and  relation (\ref{p=n}) from   Eq.(\ref{sum}) one obtains that in 
$\sim k_F-600$~MeV/c range
 \begin{equation}
 n^{A}_{p/n}({p}) \approx {1\over 2 x_{p/n}} a_2(A,y)\cdot n_d({p}),
 \label{highn}
 \end{equation}
 where $a_{NN}(A) \approx a_{pn}(A,y)\equiv a_2(A,y)$   and  
 the nuclear asymmetry parameter  is defined as  $y= |x_n-  x_p|$.  
 
 According to Ref.\cite{Sargsian:2012sm}, for the situation in which the asymmetry parameter can be considered small, ($y< 1$),  the 
 NN correlation factor $a_2(A,y)\approx a_2(A,0)$ which is a slowly changing function of  nuclear mass for $A>4$. 
 This allows us to formulate  the  {\em second} property of the high-momentum distribution of nucleons:  that,  according to Eq.(\ref{highn}), {\em the probability of a proton or neutron being in high-momentum NN correlation is inversely proportional to their  relative fractions ($x_p$ or $x_n$)  in the nucleus}.

We check the validity of  this relation by comparing the momentum distribution in the NN SRC domain  based on 
Eq.(\ref{highn}) with the realistic distributions  presented in Fig.\ref{He3_Be10}.
For this we use the estimates of $a_2$  for $^3He$ and 
$^{10}Be$ from  Refs.~\cite{McGauley:2011qc,fomin2012} and  the deuteron momentum distribution $n_d$ calculated using 
Argonne V18 NN potential~\cite{Wiringa:1994wb}, which were also used to calculate the realistic momentum distributions for $^3He$ and $^{10}Be$.
As can be seen from  these comparisons, Eq.(\ref{highn}) works rather well starting at 200~MeV/c and surprisingly for up to the 
momenta $\sim 1$~GeV/c, indicating  that the 3N SRCs are parametrically small for all momenta (as 
discussed in Sec.\ref{emergenceSRC}).

\subsubsection{Momentum Sharing in Asymmetric Nuclei}

The important implication of  the {\em second property}  is that the relative number 
of high momentum  protons and neutrons becomes  increasingly  {\em unbalanced}  with an  increase 
of the nuclear  asymmetry,  $y$. To quantify this prediction, using  Eq.(\ref{highn}) one  calculates the   fraction 
of the   nucleons having  momenta $\ge k_{F}$ as:
\begin{equation}
P_{p/n}(A,y) \approx  {1\over 2 x_{p/n}} a_2(A,y)\int\limits_{k_F}^{\infty} n_d(p) d^3p,
\label{fraction}
\end{equation}
where we extend the upper limit of integration to infinity because of 
the smaller overall contribution  from the $\ge 600$~MeV/c region.
The results of  the calculation  of $P_{p/n}$s  for medium to heavy nuclei, 
using the  estimates of $a_2(A,y)$ from  Ref.~\cite{frankfurt93,egiyan03,egiyan06,McGauley:2011qc,fomin2012} 
and $k_{F}$ from Ref.~\cite{Moniz:1971mt} are given in Table.\ref{table1}. 

\begin{table}[t]
\caption{Fractions of high momentum protons and neutrons in nuclei A.}
\vspace{0.2cm}
\centering
\begin{tabular}{llllll}
\hline\hline 
A & $P_p(\%)$  & $P_n(\%)$ & A & $P_p(\%)$  & $P_n(\%)$ \\  [0.5ex] 
 \hline
 %hayr mer
12  & 20 & 20 &
56  & 27 & 23 \\
27  & 24 & 22 &
197 & 31 & 20 \\
\hline
\end{tabular}
\label{table1}
\end{table}
The estimates in  Table.\ref{table1} indicate that  with  the increase of asymmetry $y$, the imbalance between the high-momentum 
fractions of proton and neutron grows.  For gold,  the fraction of high-momentum protons exceeds that of the neutrons by as much as 
50\%.  
 
\begin{table}[ht]
\caption{Kinetic energies (in MeV) of protons and neutrons across several nuclei} 
\vspace{0.2cm}
\centering 
\begin{tabular}{l l l l c }
\hline\hline 
A & y & $E_{kin}^p$  &   $E_{kin}^n$  &   $E_{kin}^p-E_{kin}^n$\\  [0.0ex] 
\hline 
$^{8}$He     \    & 0.50  \  & 30.13 \  &  18.60 \  & 11.53     \\ 
$^{6}$He     \    & 0.33  \  & 27.66 \  & 19.06  \  &  8.60  \\ 
$^{9}$Li       \    & 0.33  \  & 31.39 \  &  24.91 \  & 6.48    \\ 
$^{3}$He     \    & 0.33  \  & 14.71 \  & 19.35  \  &  -4.64   \\ 
$^{3}$H       \    & 0.33  \  & 19.61   \  &  14.96 \  &   4.65   \\ 
$^{8}$Li       \    & 0.25  \  & 28.95 \  &  23.98 \  & 4.97   \\ 
$^{10}$Be   \    & 0.2    \   & 30.20 \  &  25.95 \  & 4.25    \\
$^{7}$Li       \    & 0.14  \  & 26.88 \  &  24.54 \  & 2.34   \\ 
$^{9}$Be     \    & 0.11  \   & 29.82 \  &  27.09 \  & 2.73    \\
$^{11}$B     \    & 0.09   \  \  \ & 33.40 \  \  \ &  31.75 \ \ \  &   1.65    \\ 
 \hline 
\end{tabular}
\label{table2} 
\end{table}

Another implication of Eq.(\ref{fraction}) is that due to the larger relative fraction of high momentum  the minority component  
should be more energetic in asymmetric nuclei  than the majority component.  Namely, one expects a more energetic 
neutron than proton in $^3He$  and  the opposite result for neutron rich nuclei.   This expectation is confirmed 
by ab-initio calculation of  $p-$ and $n-$ kinetic energies for all nuclei (currently up to $A\le 11$)
(see Table \ref{table2} and Ref.~\cite{Sargsian:2012sm} for more details).

\subsection{Three Nucleon Correlations}
In the previous discussion, we defined a nucleon to be in a 2N SRC if its momentum exceeds $k_{F}$ and almost entirily  compensated by the 
momentum of the correlated nucleon in the nucleus. 
For a nucleon to be in a 3N SRC we assume, again,  that its momentum significantly exceeds $k_{F}$ but in this case  is balanced by two 
correlated nucleons with momenta $> k_F$. In both cases the center of mass momentum of the SRC, $p_{cm} \le k_F$. 
 
In principle, the complete nuclear wave function should contain the above-described property of 2N and 3N SRCs.
 Unfortunately, the calculation of such wave functions from  first principles is currently impossible due to poorly understood
 strong interaction dynamics at short nuclear distances as well as relativistic effects that become increasingly important at 
 large momenta of nucleons involved in short range correlations.
 However, recent theoretical studies have provided a sufficient roadmap for meaningful experimental exploration of 
 3N SRCs.

One result of such studies is  that the irreducible three-nucleon forces contribute to 3N SRCs only at very large  nuclear excitation energies of
 $>  2(\sqrt{p^2 +m_N^2} - m_N)$, $p\gtorder 700$~MeV/c, and are otherwise negligible~\cite{sargsian05}.  Thus, outside of such kinematic 
 regions, the dynamics of 3N SRCs are defined by two successive  short-range $NN$ interactions~\cite{Artiles:2016akj}.
 Such a situation highlights the difficulties of experimental identification of 3N SRCs. From the point of view of 
 extracting the nuclear spectral function, the separation of 3N SRCs are problematic
 since the expected enhancement in recoil energy distribution at  $E_{rec}\approx p^2/4m_N$ is very broad without a discernible 
 maximum~\cite{sargsian05}.  With regard to the momentum distribution, as the 3N SRCs are subleading compared to  2N SRCs~\cite{Frankfurt:2008zv} (see also Sec.\ref{emergenceSRC}),  they are parametrically small for all $p$  making the momentum distribution $n_A(p)$  rather insensitive to 3N SRCs.
  
Hence, one of the problems in experimental identification of 3N  SRCs is a proper identification of the variables 
that can unambiguously discriminate 3N and 2N  SRCs.  To this end, the relevant variable is 
the light-cone momentum fraction, $\alpha_i$  defined in Eq.(\ref{alphai}).  Due to the short-range nature of 
nuclear forces the  condition
\begin{equation}
j-1 < \alpha_i < j
\label{srcon}
\end{equation}
will ensure that scattering from $(j\times N)$-SRC is being probed~\cite{frankfurt81,Frankfurt:2008zv}. Thus, one expects that 
the extraction of $\rho_A(\alpha_i)$ at $\alpha_i>2$ will ensure the dominance of 3N SRCs.

\begin{figure}
\centering
\parbox{6cm}{
\includegraphics[width=5cm]{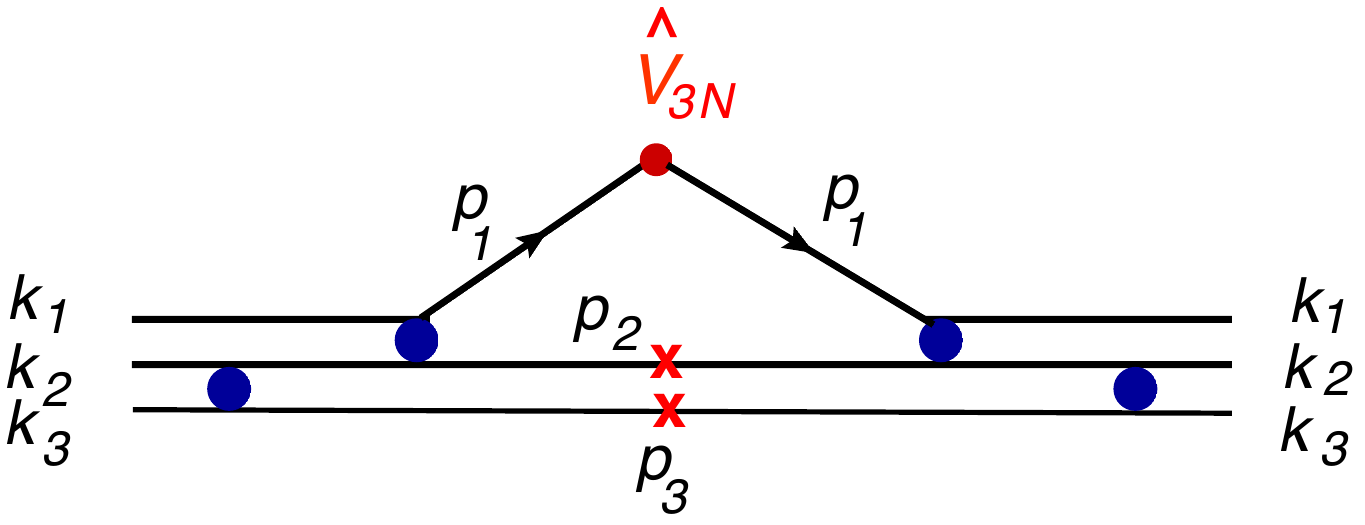}
}
\qquad
\begin{minipage}{6cm}
\includegraphics[width=6cm]{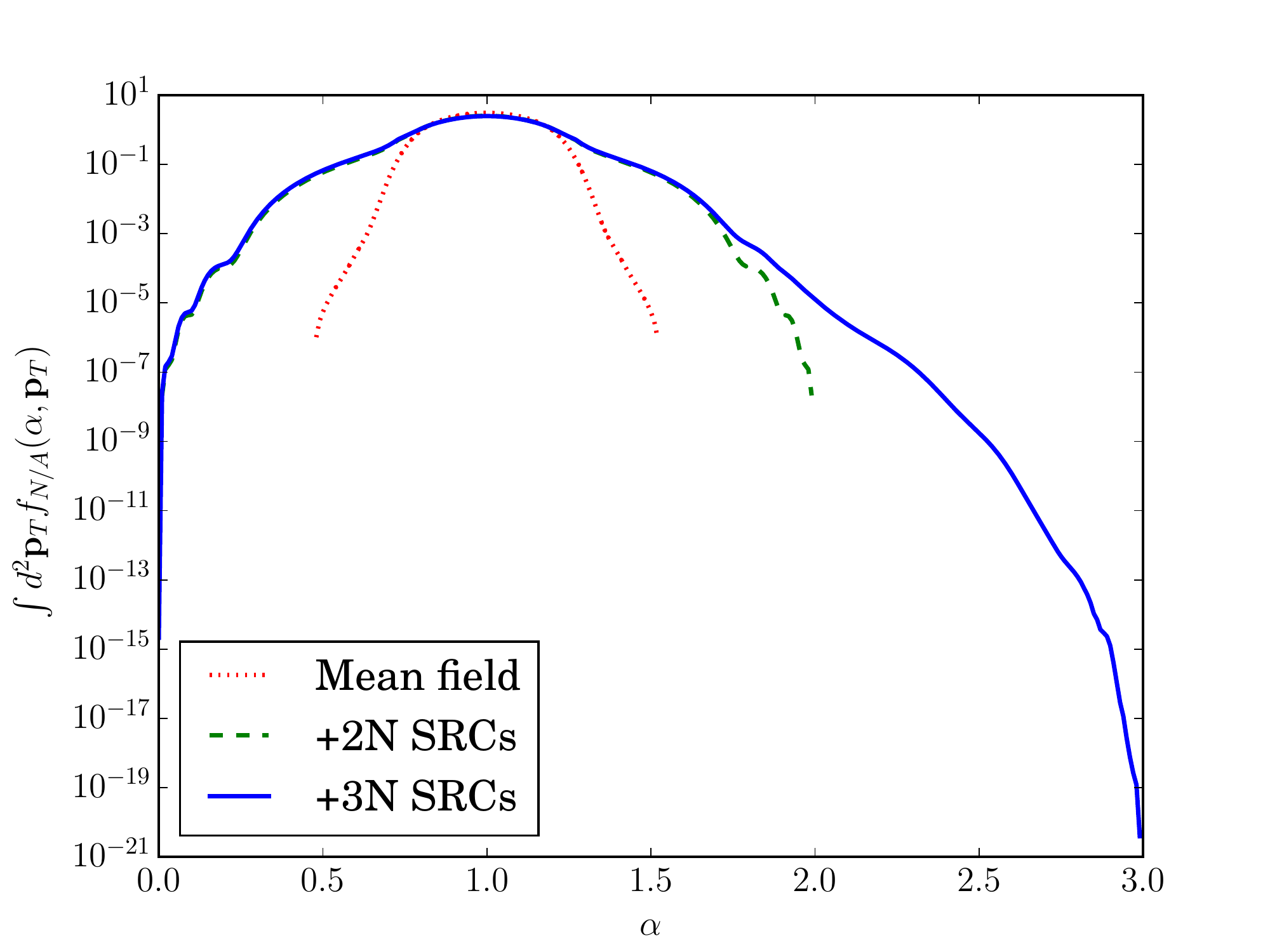}
\end{minipage}
\caption{(left panel) The 3N SRCs due to sequential 2N short range correlations. (right panel) The $\alpha$ distribution of 
the light-front density matrix. Dotted - lines - nuclear mean field contribution only, dashed  - mean-field and 2N SRC cotribution,
solid - mean field, 2N- and 3N SRCs included.  Figure adapted from Ref.\cite{Freese:2014zda}.}
\label{fig:3N_SRC}
\end{figure}

\noindent{\bf Dynamics of the 3N SRCs:} In light of the recent observation of strong dominance of $pn$ pairs in 2N SRCs~(Sec.\ref{sec_pn_dom}) for momenta of $250-600$~MeV/c and the expectation (discussed above) that 3N SRCs are predominantly due to two successive 2N short-range interactions, 
one predicts a strong implication of the $pn$ dominance in 3N SRCs as well. 
This means that  3N SRCs should predominantly have a $ppn$ or $nnp$ composition with $ppp$ and $nnn$  configurations being strongly suppressed. The diagram that will represent the 
light-cone density matrix of 3N SRCs is given in Fig.\ref{fig:3N_SRC}(left panel), where three nucleons are either in  $ppn$ or $nnp$  configuration.
The light-front spectral function  according to the diagram of Fig.\ref{fig:3N_SRC}(left panel), calculated within effective Feynman diagrammatic 
approach~\cite{Artiles:2016akj,Freese:2014zda} is expressed through the light-front density functions of NN SRCs  as follows: 
\begin{eqnarray}
P^N_{A,3N}(\alpha_1,p_{1,\perp},\tilde M_N)  &  =  & \int {3-\alpha_3\over 2(2-\alpha_3)^2} \rho_{NN}(\beta_3,p_{3\perp}) 
\rho_{NN}(\beta_1,\tilde k_{1\perp}) 
2 \delta(\alpha_1 + \alpha_2  + \alpha_3 - 3)\nonumber \\
& &  \delta^2(p_{1\perp} + p_{2\perp} + p_{3\perp}) 
 \delta(\tilde M_N^2 - M_N^{3N,2})  d\alpha_2 d^2 p_{2\perp}d\alpha_3d^2 p_{3\perp},
 \label{P3Nsrc}
\end{eqnarray}
where within colinear approximation $\beta_3 = \alpha_3$, $\beta_1 = {2\alpha_2\over 3-\alpha_3}$ as well as $\tilde k_{1\perp} = p_{1\perp} + {\beta_1\over 2} p_{3\perp}$ with 
$\alpha_i$ representing the light-cone momentum fraction of 3N SRCs carried by the correlated nucleon $i$.  The 3N SRC spectral function uses as  an input the density functions of 2N SRCs,  $\rho_{NN}$, which can be extracted from studies of two-nucleon correlations.

The dynamics of 3N SRCs, according to the correlation scenario of Fig.\ref{fig:3N_SRC}(left panel), have several unique features that can be verified experimentally.  One is that the per nucleon probability of finding a nucleon in a  3N SRC, $a_{3N}$, should be  proportional to the square of the probabilities of  
2N SRCs,  $a_{2N}$, i.e. 
\vspace{-0.2cm}
\begin{equation}
a_{3N} \sim a_{2N}^2.
\vspace{-0.1cm}
\label{a3sima2a2}
\end{equation}
Another  feature is that the dominant 3N SRC configuration will be one in which the two nucleon spectator system has a 
minimal mass $m_S\gtorder 2m_N$,  corresponding to small relative momentum in the recoil NN system, $k_{NN} = {\sqrt{m_S^2 - 4m_N^2}\over 2}$. The 
condition $k_{NN}\ll m_N$  and the fact that 
isotriplet two-nucleon states with low relative momenta are strongly suppressed~\cite{Sargsian:2004tz} compared to the isosinglet states  results in a 
strong dependince of the 3N SRCs on the isospin structure of NN recoil system.  Namely,  3N SRCs are dominated by configurations in which the recoil two-nucleon system is in  the isosinglet state.  

We conclude the discussion on 3N SRCs illustrating a calculation of $p_T$ integrated light-front density function~\cite{Freese:2014zda}  in 
Fig.\ref{fig:3N_SRC} (right panel) containing mean-field, 2N and 3N SRC contributions. As the figure shows, despite the 3N SRCs being subleading to the 2N SRCs, they are well separated in $\alpha$, as was discussed above.

\subsubsection{Non-Nucleonic Degrees of Freedom and Medium Modifications} 
 Recent experimental and theoretical studies of SRCs have advanced our understanding of the domain of 
 two-nucleon correlations for momenta up to $600$~MeV/c.   This provides important groundwork for 
the  extension of  SRC studies to the more unchartered territory of hadron-quark transition.  
The studies of isospin structure of NN SRCs~\cite{piasetzky06,shneor07,subedi08} found not only the $pn$ dominance but also observed 
that the sum of the absolute $pp$ and $pn$ probabilities are  close to unity, indicating that practically all nucleons in 
this momentum range (with accuracy in the order of  {\em few}\%) belong  to two nucleon SRC.  This is an important 
observation since with a high level of confidence it excludes the explicit non-nucleonic degrees of freedom in 
this momentum range.

By contrast, nucleons are composite particles, and therefore  inter-nucleon interactions  
in SRCs should lead to a deformation of the  bound nucleon wave function.  
Such a modification of the bound nucleon is manifested in the  experimental observation of the difference between 
partonic distributions of the   free  and bound nucleons, commonly referred as EMC 
effect~\cite{Aubert:1981gw}.  The most recent EMC effect measurement was performed for the wide range of 
nuclear targets, focussing on light nuclei~\cite{Seely:2009gt}, with  the interesting observation that the extent of the modification is not proportional 
to the average density of nuclei but rather to the local density at which the bound nucleon is probed.
The last observation can possibly indicate the important role that SRCs play in the 
medium modification of the partonic distribution of nucleons in the nuclei. This expectation is 
further enhanced with the observation of apparent correlation between the strengths of  
the EMC effect and  2N SRCs~\cite{Weinstein:2010rt,Arrington:2012ax}. Similar, apparent correlations~\cite{Arrington:2012ax} have been 
observed also between EMC effects and nuclear excitation energies all indicating the enhancing role of SRCs in 
the modification of the internal structure of bound nucleons.

\subsection{Nuclear Dynamics at Core Distances}

One of the most fascinating aspects of  nuclear physics at short distances is the repulsive core. 
With the  coming of age of high energy and high intensity accelerators such as CEBAF12, LHC and JPARC as well as currently 
discussed  electron ion collider~(EIC) 
there is a real possibility for systematic  exploration  of nuclear dynamics at core distances.  We will characterize core distances as those where the probed relative distance is smaller than the radius of the nucleon which will mean substantial
overlap of two nucleons in the nuclear medium.

Currently very little is known about the structure and the dynamics of nuclear repulsion at core distances.  The modern NN potentials 
employ the Wood-Saxon ansatz  while models based on  one boson-exchanges introduce vector-meson exchanges to reproduce repulsion at distances  smaller than the size of the exchanged mesons. QCD-based models employ other extreme scenarios in which two nucleons 
 are collapsed into a six-quark state. 

The relative momenta in the NN system which will be sensitive to the core dynamics can be estimated based on the threshold of inelastic $N\to  \Delta$ transition:
\begin{equation}
\sqrt{M_N^2 + p^2} - M_N \ge M_{\Delta}-M_N
\end{equation}
which results in $p\ge 800$~MeV/c.   
Thus one of the direct ways of reaching the core is to probe internal momenta 
in $\sim 1$~GeV/c region.

\subsubsection{Super-Fast Quarks in Nuclear Medium}
\label{sfquarks}
Experimentally, it is possible to probe internal momenta in nuclei in the range of $\ge 1$~GeV/c by considering deep inelastic scattering 
from superfast quarks in nuclei~\cite{frankfurt88,Sargsian:2002wc}.

We define superfast quark as partons that have been probed via  nuclear DIS scattering at  Bjorken $x>1$.   Since in an isolated nucleon 
at rest partons cannot carry momentum fractions  larger than one, DIS scattering at $x>1$ is possible if partons in nearby nucleons are sharing momentum.  This requires these two nucleons to be in very close proximity.

There are two basic reasons that superfast quarks probe  high-density 
fluctuations in the nuclear medium.  The first is kinematical.
The initial longitudinal momentum of a nucleon probed 
in DIS scattering is~\cite{Sargsian:2002wc}:
\begin{equation}
p_{i}^z = m(1-x_- x\left[{W^2-m^2\over Q^2}\right]),
\label{pin}
\end{equation}
where $W$ is the final mass produced on the nucleon in the nucleus. 
As Fig.\ref{fig:SRCkin}(left panel) shows, in the DIS region at $W\ge 2.5$~GeV, the virtual photon will probe 
a nucleon  with momentum  $p_{i}^z\sim -1~GeV/c$. 

The second reason that superfast quark probes very high-momentum 
component in the nuclear medium is dynamic. Due to  QCD  evolution, the parton at $x$ is evolved from the original partons 
with momentum fractions $x_0\ge x$ which increases with $Q^2$.  Thus, the alternative 
way of probing short space-time separations in the nuclear medium will 
be a measurement with fixed $x$ and increasingly large $Q^2$.
Such studies will probe QCD dynamics in the extreme nuclear conditions with the potential of opening up uncharted
territory for nuclear QCD.  

The extreme nuclear  dynamics may include multi-nucleon short range correlations~\cite{frankfurt88,Sargsian:2002wc,Freese:2014zda}, 
explicit quark degrees of freedom such as 6-quark clusters~\cite{Carlson:1994ga}, or single-quark  momentum exchanges between strongly
correlated nucleons~\cite{Sargsian:2007gd}.   

Superfast quarks can be probed in different processes including inclusive DIS from nuclei 
at $x>1$~\cite{frankfurt88,Sargsian:2002wc,Sargsian:2007gd}, the processes such as semi-inclusive nuclear DIS  
with tagged spectator nucleons~\cite{Melnitchouk:1996vp,Cosyn:2010ux},  DIS production in the forward direction  
with $x_{F}>1$ or  large transverse momentum dijet production  in  $p + A \rightarrow \mathrm{dijet} + X$ reactions  at 
LHC kinematics~\cite{Freese:2014zda}.

\section{Recent Advances in Experiment}

The beginning of Jefferson Lab's operation in 1990's  started a truly new era in 
exploration of the nuclear structure at short distances. The SRC  program included 
high $Q^2$ ($\ge 1$~GeV$^2$) inclusive $A(e,e')X$  measuremens\cite{Arrington:1995hs,Arrington:1998ps,Arrington:2001ni,egiyan03,egiyan06,fomin2012}
complemented  with semi-inclusive triple-coincidence\cite{shneor07,subedi08,Korover:2014dma,Hen:2014nza} and exclusive deuteron electrodisintegration\cite{Egiyan:2007qj,Boeglin:2011mt} 
experiments.

The most important results on NN SRCs have been obtained recently from inclusive and triple-coincidence measurements which we will review below.

\subsection{Inclusive high $Q^2$ Processes at $x>1$.}

The main motivation in any experiment aimed at studying  SRCs is to probe the bound nucleon with momentum exceeding characteristic 
Fermi momentum $k_{F}\sim 250$~MeV/c.
As it was discussed in Sec.\ref{sec2.2}, the more relevant quantity in this respect is light-front momentum fraction $\alpha_i$ (Eq.(\ref{alphai})) for which, choosing condition of Eq.(\ref{srcon}) 
will allow to probe $j$-nucleon short range correlations.  
The technical question for a given experiment is how to determine the parameter $\alpha_i$ for different regions of NN correlation dominance.

One way to achieve such a measurement is in the quasileastic nuclear reactions in which from the condition $(p_i + q)^2 = M_N^2$ one obtains:
\begin{equation}
\alpha_i = x\left(1 + {2p_{i,z}\over q_0 + |q|}\right) +
{M_N^2 - m_i^2\over 2m_N q_0} ~,
\label{incl_kin}
\end{equation}
where $p_{i,z}$ is the longitudinal momentum of the initial nucleon
along the direction of the transferred momentum $q$ in the nuclear rest frame.  
From the above equation  one observes that in the asymptotic limit of large $Q^2$,  $\alpha_i \approx x$ and thus 
choosing $x>1$ allows to satisfy conditions necessary to probe multi-nucleon correlations.  Since Bjorken $x$ is defined 
by kinematics of electron scattering only, $x = {Q^2\over 2M_N q_0}$, this consideration indicates that NN correlations can 
be probed in inclusive $A(e,e^\prime)X$ experiments.

\begin{figure}
\centering
\parbox{6cm}{
\includegraphics[width=6cm]{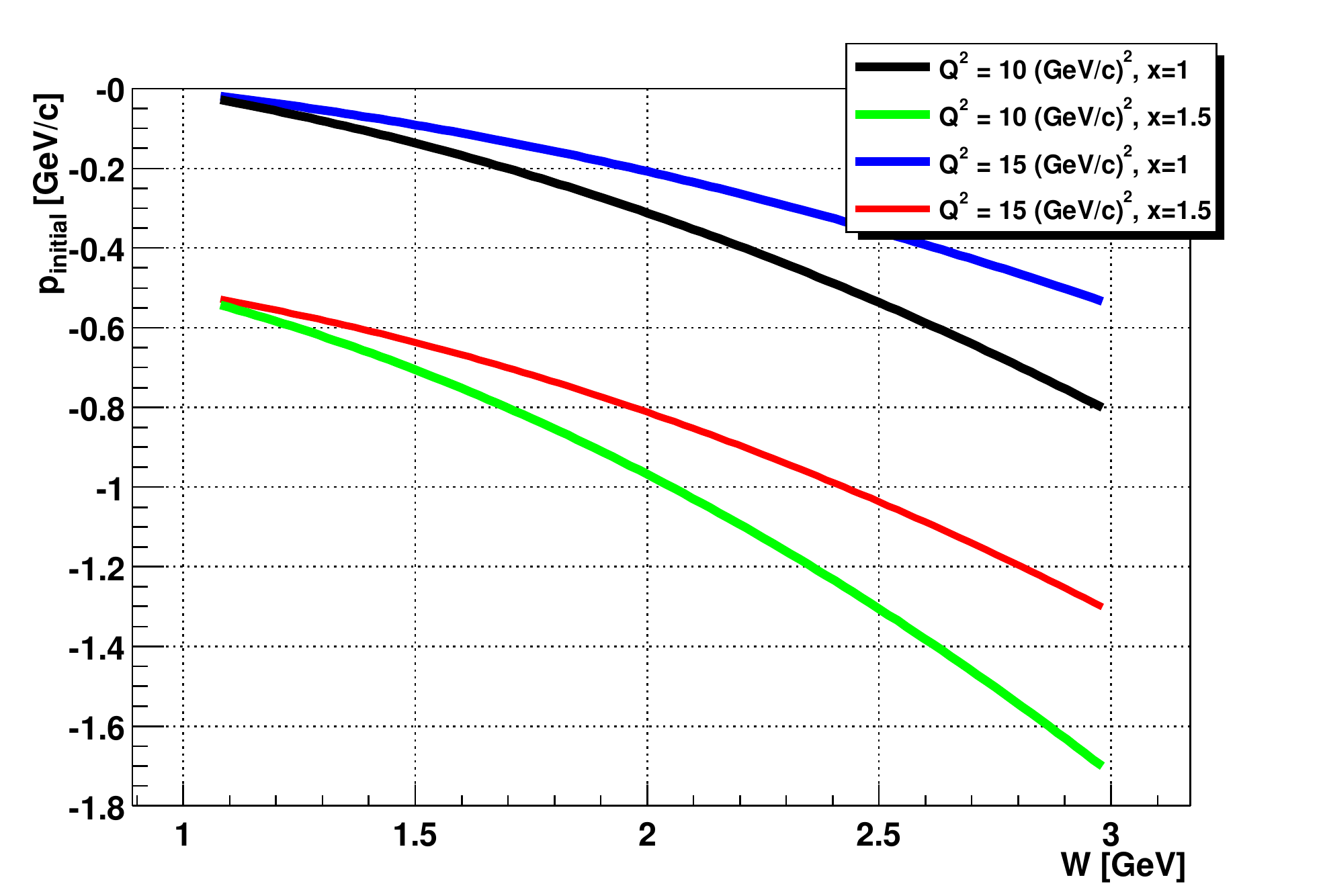}
%\caption{First.}
%\label{fig:2figsA}
}
\qquad
\begin{minipage}{6cm}
\includegraphics[width=4.5cm,angle=270]{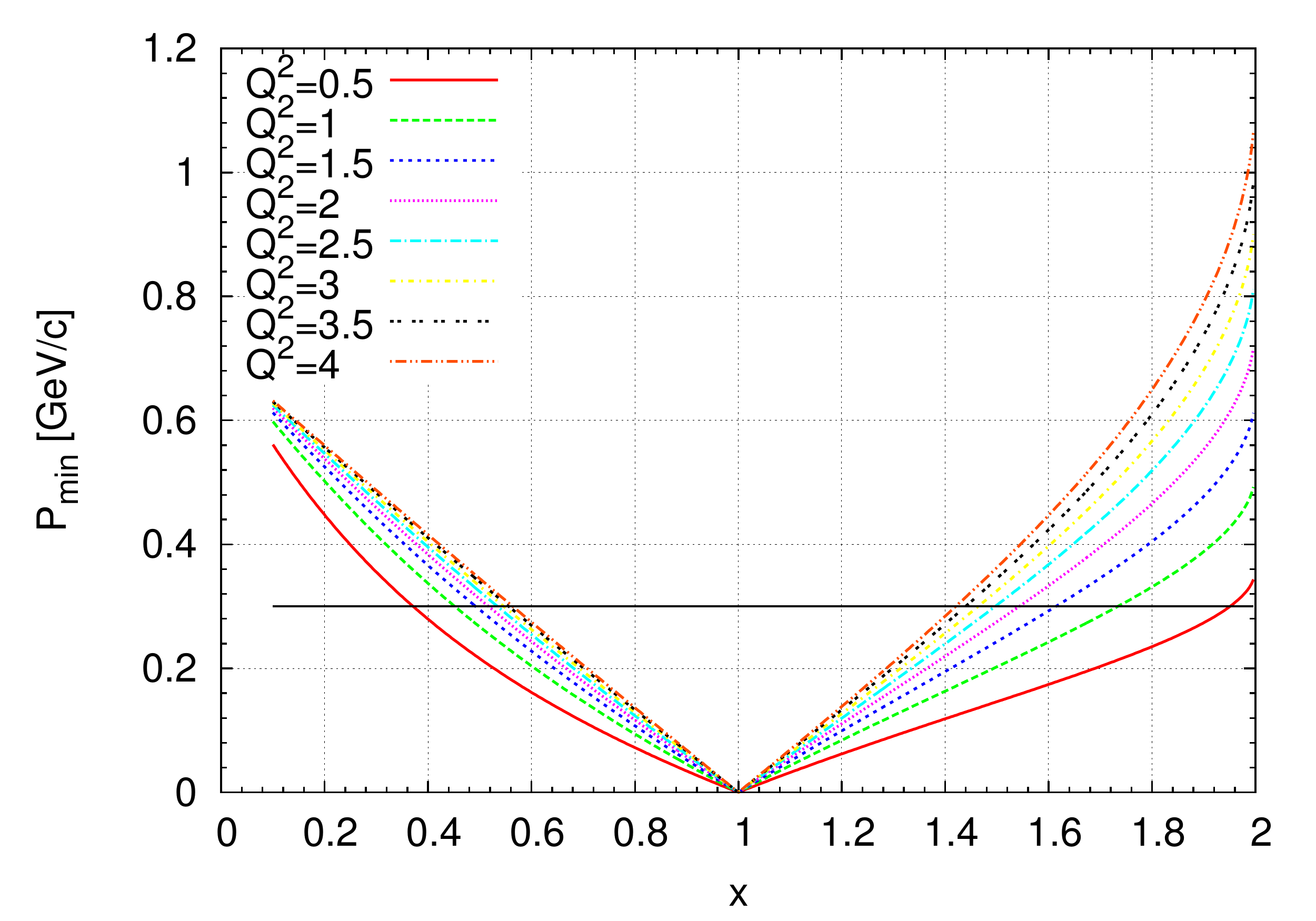}
%\caption{Second.}
%\label{fig:2figsB}
\end{minipage}
\vspace{-0.6cm}
\caption{(left panel)  Dependence of $p_{in}^z$ on the final mass $W$ produced in  DIS scattering off the bound nucleon in the nucleus.
(right panel) The minimum momentum for quasi-elastic $\gamma^* + 2N \rightarrow N + N$ scattering  as a function of $x$   
for different values of $Q^2$. }
\label{fig:SRCkin}
\end{figure}

\subsubsection{Probing 2N SRCs in $1 < x < 2$ Region}
In the above discussion we considered the asymptotically large $Q^2$ to illustrate the simple relation between $\alpha_i$ and Bjorken $x$. 
%However even
At finite  values of  $Q^2$  one can use  Eq.(\ref{incl_kin}) to identify the optimal kinematics for 
separation of 2N correlations.   For this we notice that Eq.(\ref{incl_kin}) can be solved for the longitudinal component of $p_i$ that  
represents the minimal magnitude for the  momentum of the bound nucleon. Then one can find the suitable $x,Q^2$ kinematics for which 
 $|p_{min}|> k_{Fermi}$ to isolate 2N SRCs. Fig.~\ref{fig:SRCkin}(right panel) shows the relationship between $p_{min}$, $x$ and $Q^2$, leading one to conclude that to observe 2N SRCs at $x \ge $1.4, a  $Q^2=1.4$~GeV$^2$ is required. The figure also shows that one will not be able to probe correlations at $Q^2 \le 1$~GeV$^2$.  The existence of such a threshold in $Q^2$ was experimentally observed in the dedicated measurement of inclusive cross sections at $x>1$ in Refs.\cite{egiyan03}.

The  minimal initial momentum of the bound nucleon can be used to estimate~\cite{frankfurt93} the  corresponding 
light-cone momentum fraction, $\alpha_{2N}$  which is defined by the parameters of 
scattered electron only:
\begin{equation}
\alpha_{2N} = 2 - \frac{q_- + 2m_N}{2m_N}
\Bigg( 1+\frac{\sqrt{W^2-m_N^2}}{W^2} \Bigg) ~,
\label{eq:alpha2n}
\end{equation}
where $W^2 = 4m_N^2 + 4m_N \nu - Q^2$.  
 The advantage of discussing the inclusive cross section based on $\alpha_{2N}$ 
representation is that in the high energy limit one expects that inclusive cross section to  factorize into the product of 
cross section of electron-bound nucleon scattering  and the light-front density matrix of the nucleus $\rho_A(\alpha_{2N})$.  
Therefore the inclusive scattering in principle allows an extraction of $\rho_A(\alpha_{2N})$ from the measured cross sections.

The first indication of the onset of 2N SRC regime in inclusive scattering is the appearance of the plateau in the ratios of 
the cross sections measured for nuclei $A$ to the deuteron\cite{fomin2012}  or $^3He$\cite{egiyan03,egiyan06} in the $x>1$ region  at
 large $Q^2$.  This plateau is the result of the nuclear high momentum's factorization in the form of Eq.(\ref{highn}). 
 
\begin{figure}[h!]
\vspace{-3cm}
\hspace{-2cm}
\includegraphics[width=0.55\textwidth, angle=270,trim={0cm 2cm 0 5cm}, clip]{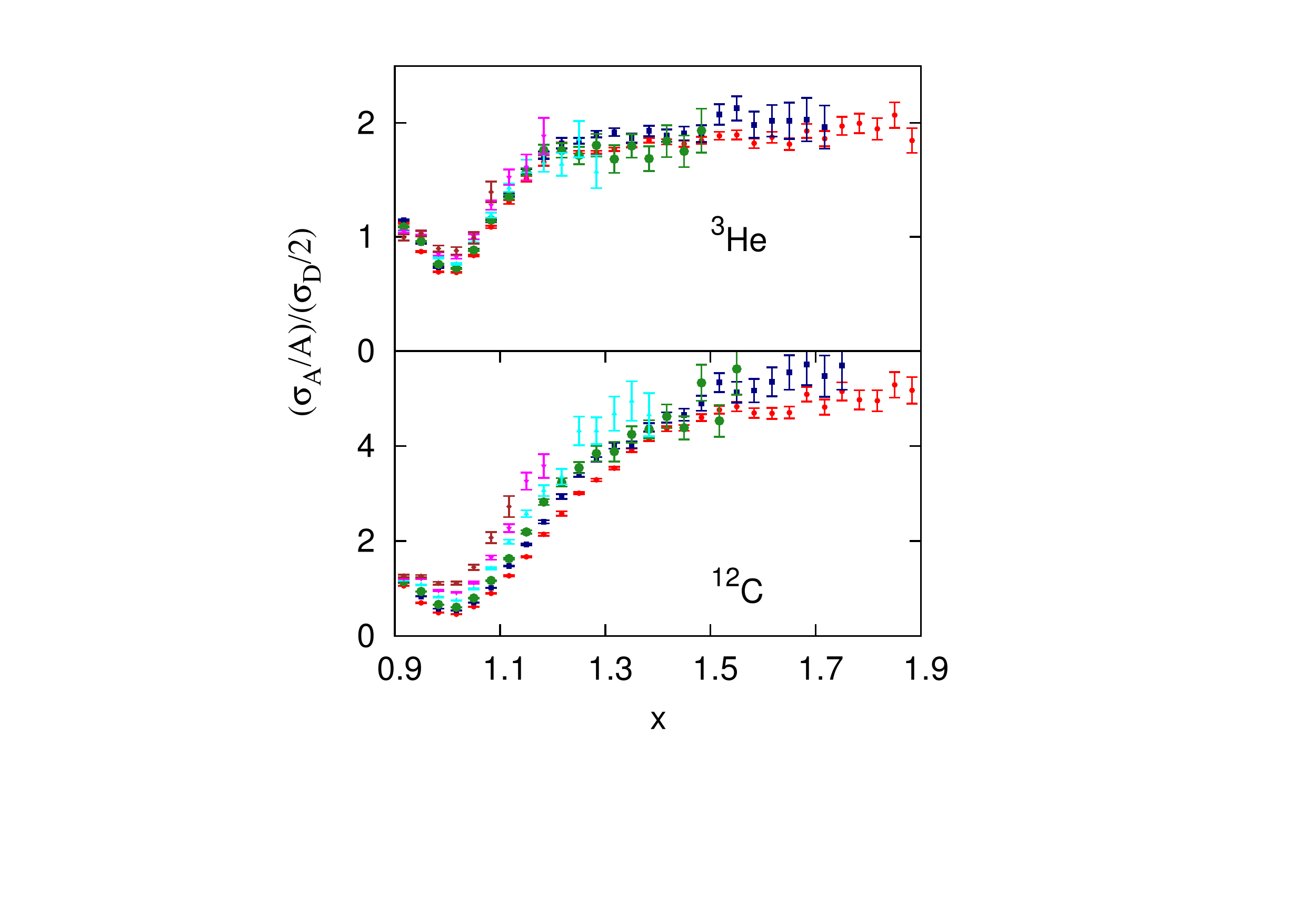}
%\vspace{-4cm}
%\hspace{4cm}
\includegraphics[width=0.55\textwidth, angle=270, trim={0cm 2cm 0 5cm},clip]{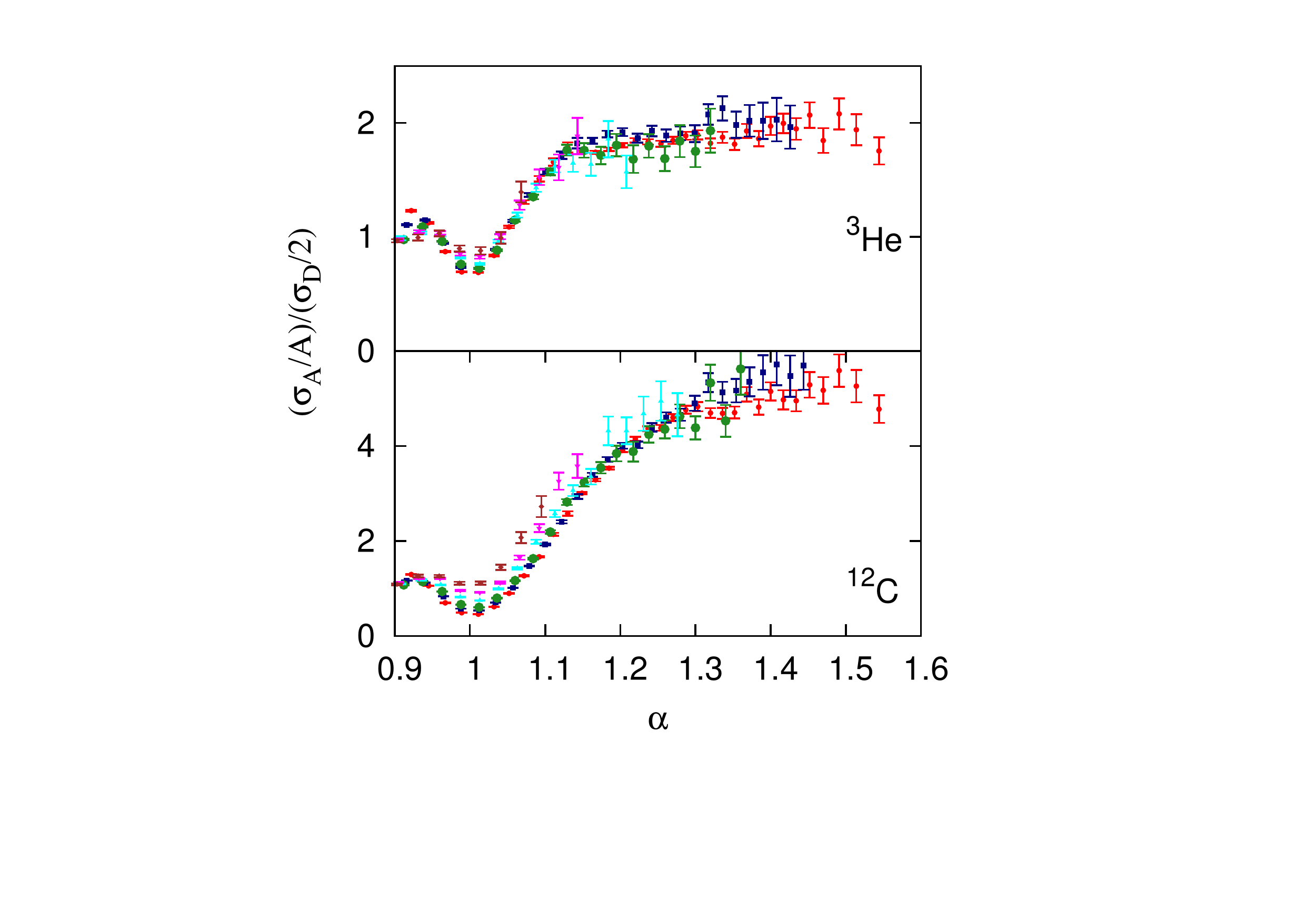}
\vspace{-4cm}
\caption{$\sigma_A/\sigma_D$ vs $x$ (left panel) and $\alpha$ (right panel) for selected targets from JLab E02-019 experiment.  See text for details.}
\label{fig:ratios_2n}
\end{figure}

The recent results of  such measurements are presented in  Fig.\ref{fig:ratios_2n}(left-hand side, plotted vs $x$) for $^3He/D$ and $^{12}C/D$ ratios.  The different sets of colored points correspond to data taken at different Q$^2$ values, ranging from 2.5~GeV$^2$ to 7.4~GeV$^2$ 
(as evaluated at $x=1$).  Only points with uncertainties under 10\% are shown, meaning that only the lowest Q$^2$ data reach the highest $x$ values.

One of the  interesting features of the $x$ dependence of the experimental ratios is that the threshold for the onset of the 
scaling is pushed out further in $x$ for $^{12}$C as compared to the $^3He$. This can be understood from the fact that for heavier nuclei, $k_{fermi}$ is higher, and the 2N SRC region begins at higher $x$ values, as the mean-field contribution persists for longer.

Another feature seen in the $x$ dependence is that for given nucleus the onset of scaling is Q$^2$-dependent.  Such $Q^2$ dependence is
more visible in the  $^{12}C/D$ ratios where one observes that the data at higher Q$^2$ values rise towards the plateau value sooner.  
Such a $Q^2$ dependence can be understood from Eq.(\ref{eq:alpha2n}), where for a given $x$ value, the $\alpha_{2N}$ values %scale
 change
 with $Q^2$.   However if we consider the $\alpha_{2N}$ distribution of the 
ratios instead of $x$ then such a $Q^2$ dependence should diminish.  
This is what is observed in the experimental data shown in the right-side panel of Fig.~\ref{fig:ratios_2n}.   This fact is one of the most important arguments for the validity of the 2N SRC picture of inclusive scattering. 

Even though the observation of the plateau is consistent with the expectation of scattering from 2N SRCs, to study the dynamics of the SRCs in more detail, one needs to address several theoretical issues. 
One issue is the shape of the ratios.  If the contribution in the $x>1$ region is only from 2N SRCs, then  
the expectation is that the shape of the  plateau, should be the same for both  $^3$He and $^{12}$C and for other nuclear targets~\cite{fomin2012}.  
The nuclear cross-sections can extend upto $x\approx A$, meaning the $D$ cross-section has to go to zero by $x=2$, 
whereas for $A>2$, strength continues well past that.  Therefore, we expect a slight rise in the $A/D$ ratios as we approach the $x=2$ limit. 
This rise due to different rates of cross-section fall-off between $A$ and $D$ gets larger with increasing nucleus size. 
Additionally, for $A>2$, the correlated pair will experience center-of-mass motion in the field of the other nucleons.  
This will distort the momentum distribution compared to that of the deuteron, enhancing the high-momentum tail region.  
Finally, $A>2$ nuclei will have contributions from 3N correlations that could be appearing near the $x\approx$ 2 region.

All of the above factors are essential for the extraction of the  parameter $a_2(A,z)$  from the cross-section ratio. 
A correction is required for the center-of-mass motion of the pair, and a cut in $\alpha$ must be placed to isolate the region that is 
dominated by 2N correlations.  The most recent analysis \cite{fomin2012} did both, with a cut of $1.5< \alpha<1.9$ and rudimentary 
center-of-mass motion calculations, yielding corrections upto 20\% for heavy nuclei.  The calculation entails comparing a calculated 
deuteron momentum distribution to one smeared with realistic center-of-mass motion~\cite{CiofidegliAtti:1995qe}.  
Better evaluations of the center-of-mass corrections are desirable for future experiments.
%ADD INFO ABOUT FUTURE EXPERIMENTS.

\subsubsection{Probing 3N SRCs in $x>2$ Region}

%\subsection{Probing 3N SRCs via inclusive scattering}
According to Eq.(\ref{srcon}) accessing the region of $\alpha_i\ge 2$ will allow us to probe 3N SRCs.  However, while in the case of the 2N SRCs one needs to simply go beyond $k_F$, for 3N SRCs, the transition from two- to three- nucleon 
SRCs is more complicated. 
Early measurements of inclusive cross-section ratios at $x>1$ relied on the idea that a second plateau, corresponding to 3N strength should 
be observed at $2.4\ltorder x\ltorder 3$ and $Q^2\ge 1.4$~GeV$^2$, analogous to the 2N SRC plateau in the  $1.5\ltorder x\ltorder 1.9$ region. 
Data from JLab's Hall B that appeared to support this observation were first published by Ref.~\cite{egiyan06} with $^4He/^3He$, $^{12}C/^3He$, 
and $^{56}Fe/^3He$ ratios.  However, a later experiment in JLab's Hall C~\cite{fomin2012} did not observe a second plateau, albeit the uncertainties 
were significantly higher, thereby not excluding the possibility.  The two sets of data showed excellent agreement in the 2N correlation region, 
as can be seen in Fig.~\ref{fig:3N_SRC}.  One possibility for this discrepancy was the different kinematics of the two measurements.  As a kinematic threshold exists for the observation of 2N correlation plateaus, a 3N analog is expected, but the Q$^2$ threshold is not as easily obtained. 
The CLAS data were taken at an average Q$^2$ value of 1.4~GeV$^2$  compared to 2.7~GeV$^2$ for the Hall C data.  
Since 1.4~GeV$^2$ is the threshold for 2N correlations, one would expect needing a higher value to observe a 3N plateau, suggesting that 
the Q$^2$ of the CLAS data may have been too low.  However, this does not explain away the apparent plateau at $x>2.25$. 
\begin{figure}
\centering
\parbox{6cm}{
\includegraphics[width=4cm,height=6.3cm,angle=270]{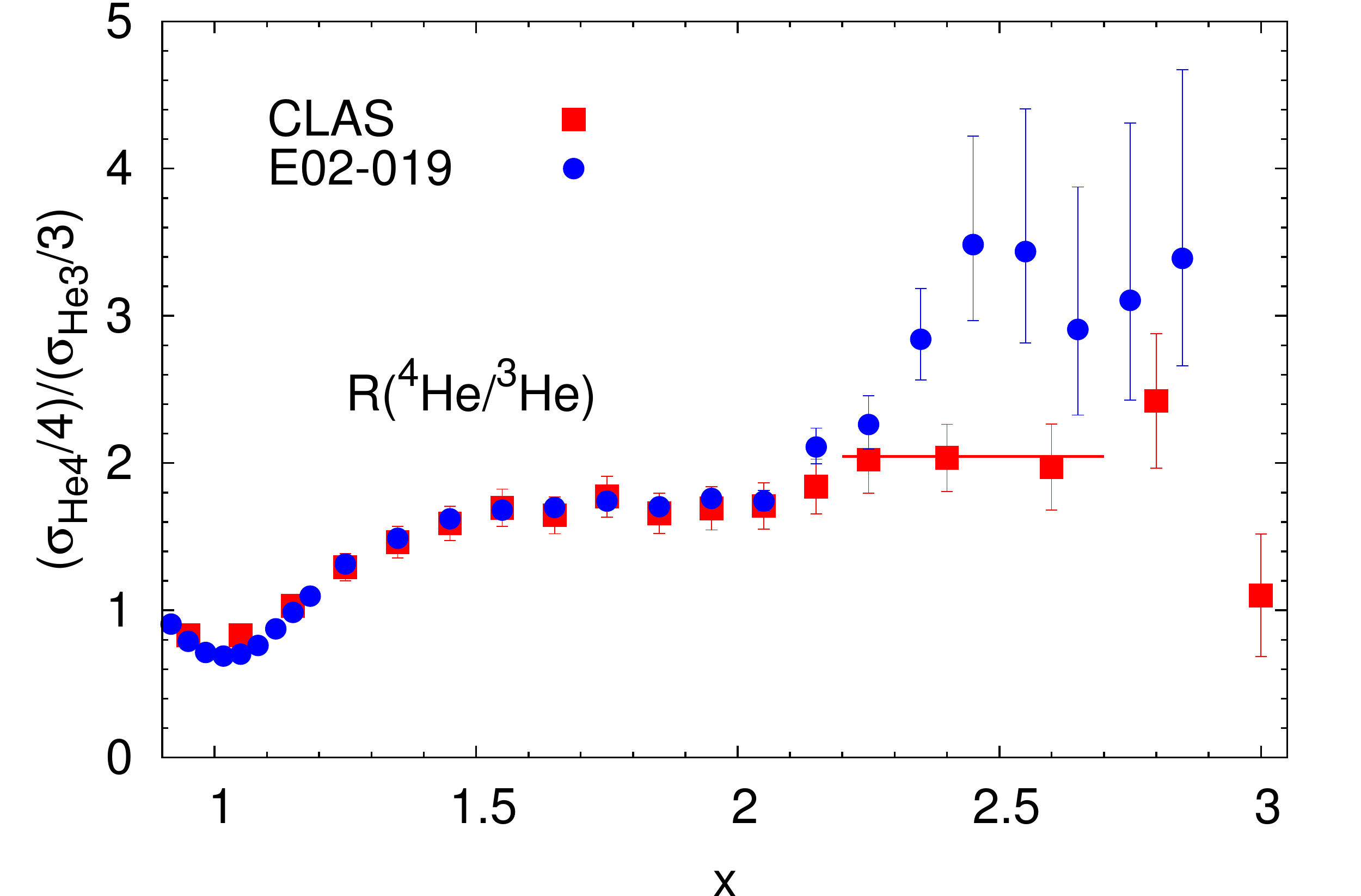}
%\caption{First.}
%\label{fig:2figsA}
}
\qquad
\begin{minipage}{6cm}
\includegraphics[width=4cm,height=6cm,angle=270]{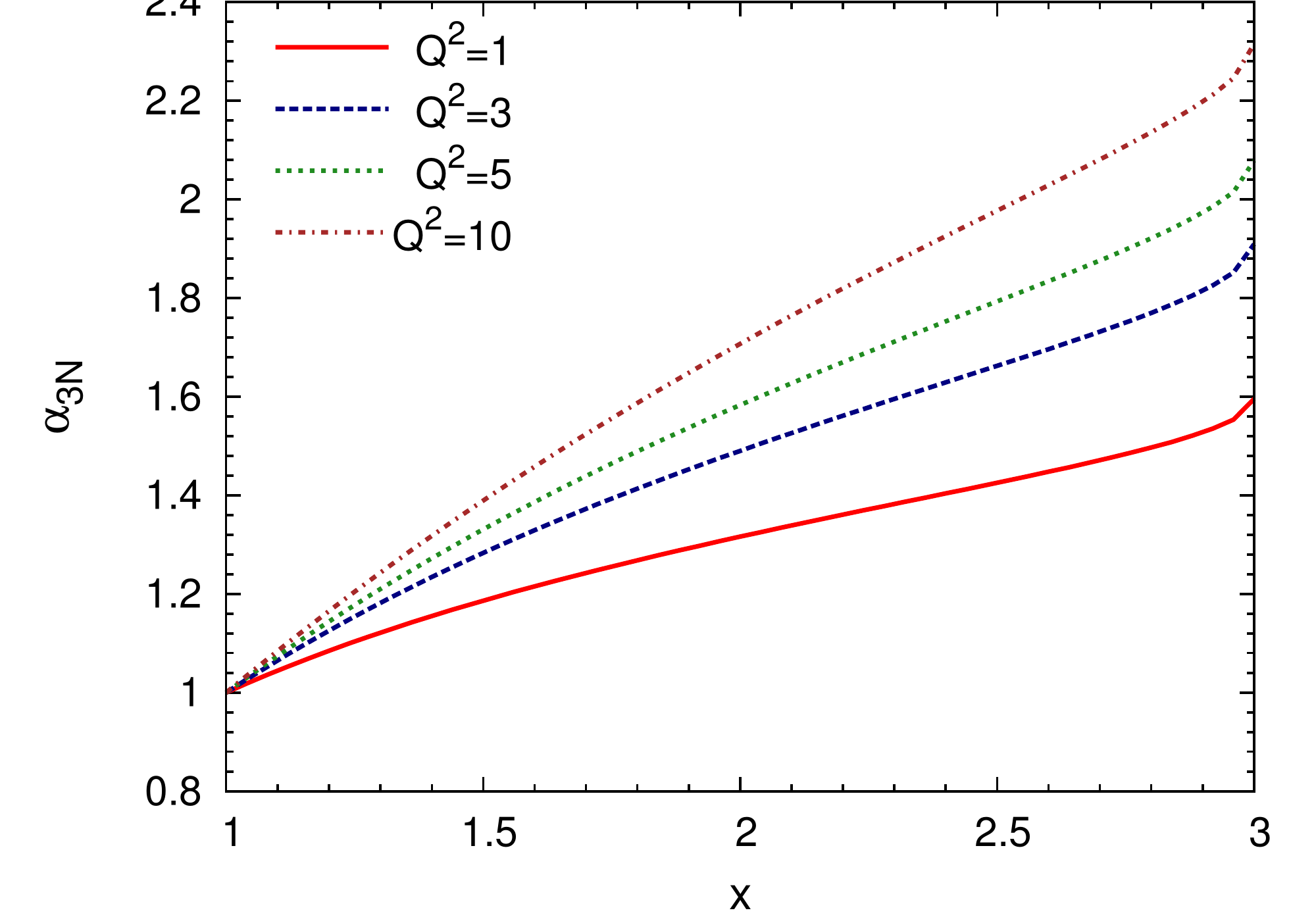}
%\caption{Second.}
%\label{fig:2figsB}
\end{minipage}
\vspace{-0.6cm}
\caption{(left panel) The ratio of $(e,e^\prime)$ cross section of $^4He$ and $^3He$ targets as a function of Bjorken $x$.
Figure adapted from Ref.\cite{fomin2012}. 
(right panel) The $\alpha_{3N}$ relation to the Bjorken $x$ for different $Q^2$.}
\label{fig:3N_SRC}
\end{figure}

A recent reanalysis of the CLAS data~\cite{Higinbotham:2014xna} shows that this  plateau can be the effect of bin migration. 
Both experiments measured cross sections as a function of the scattered energy of the electron ($E'$) and converted to $x$. 
The resolution of the CLAS spectrometer is almost an order of magnitude lower than than of the High Momentum Spectrometer in Hall C, 
and at large $x$ values, $x\ge 2$, it is larger than the size of the bins. This effect, combined with the fact that the $^3$He cross-section 
is exponentially approaching zero in this region results in significant bin migration.  Specifically, most of the points making up the 3N SRC plateau in the CLAS data~\cite{egiyan06} came from the same $E'$ bin.

The question of why JLab E02-019, whose data were taken at higher values of Q$^2$ with a high resolution spectrometer, did not observe a 3N SRC plateau remains.   In this respect one can consider the variable $\alpha_{3N}$, a counterpart of $\alpha_{2N}$ for 2N SRCs in Eq.(\ref{eq:alpha2n}). Here, $\alpha_{3N}$ properly accounts for the mass of the 3N system and the recoil of the 2N system 
 (assuming a configuration where one high-momentum nucleon is balanced by two others). 
From theoretical analysis of hadroproduction reaction,  a minimum $\alpha$ value of 1.6 is required to isolate high-momentum nucleons born in 
3N SRC. 
%The highest Q$^2$ of existing JLab data fall slightly below 3~GeV$^2$.  
Fig.~\ref{fig:3N_SRC} (right hand side) shows that a minimum Q$^2$ of 5~GeV$^2$ is needed to access this region, whereas the existing JLab data were taken below 3~GeV$^2$.  
However $Q^2 \ge  5~GeV^2$, is within reach at JLab with the 12~GeV upgrade. 
 It also means that future analyses should be done using the $\alpha_{2n,3n}$ variables, rather than the traditional $x$, as they can unambiguously connect to 2N and 3N SRC dominance regions.

\subsubsection{Upcoming Inclusive Measurements}
JLab experiment E08-014 took data in Hall A with both of the High Resolution Spectrometers, focusing on the $x>2$ region, aiming to map out the 
onset of the 3N plateau via a Q$^2$ scan as well as taking data with $^{40}$Ca and $^{48}$Ca targets to study the isospin dependence of SRC. 
Fig.~\ref{fig:3N_SRC} (right hand side) suggests that the kinematics probed by this experiment at 6~GeV were not sufficient to see a 3N plateau, but the precision 
of the data on the calcium target should be sufficient to yield interesting results.  

JLab plans to perform additional inclusive SRC measurements are planned at JLab for the 12~GeV era.  These include high-precision measurements 
on A=3 nuclei in Hall A that can be compared to calculations.  Additionally, E12-06-105 will take data on a variety of nuclei, both light and heavy 
with scans in Q$^2$ to explore the onset of the 3N plateau as well as to study the nuclear dependence of 2N SRC.  
 The most important aspects of SRC exploration with 12 GeV energy is the possibility of  unambiguous verification of the existence of 
3N SRCs,  access to the domain of the nuclear repulsive core in the NN correlation as well as probing 
the superfast quarks in deep-inelastic inclusive processes at x>1 kinematics (see Section 4.2).

%\noindent  \fbox{\bf  4 pages }

%\input{3_2emc}

\subsection{Nucleon-Nucleon Correlations and the EMC Effect}

In 1983 the European Muon Collaboration (EMC) published their surprise
deep inelastic scattering result~\cite{Aubert:1981gw}  which showed a dip in the per nucleon cross section ratio of
heavy to light nuclei when plotted vs. Bjorken $x$: a ratio that naively one would expect to be unity
up to the Fermi motion effects.
This surprising result has been reproduced many times~\cite{Gomez:1993ri,Seely:2009gt,Adams:1994ri,Amaudruz:1995tq} 
and this dip in the cross section ratio is now commonly referred to as the EMC effect.
Many possible explanations for this unexpected phenomenological result have been put
forth over the years though no definitive solution to the EMC effect
puzzle has been agreed upon.~\cite{Geesaman:1995yd,Malace:2014uea,Higinbotham:2013hta,Hen:2013oha}.

\begin{figure}
\centering
\parbox{5.5cm}{
\includegraphics[width=5.5cm,height=5.3cm]{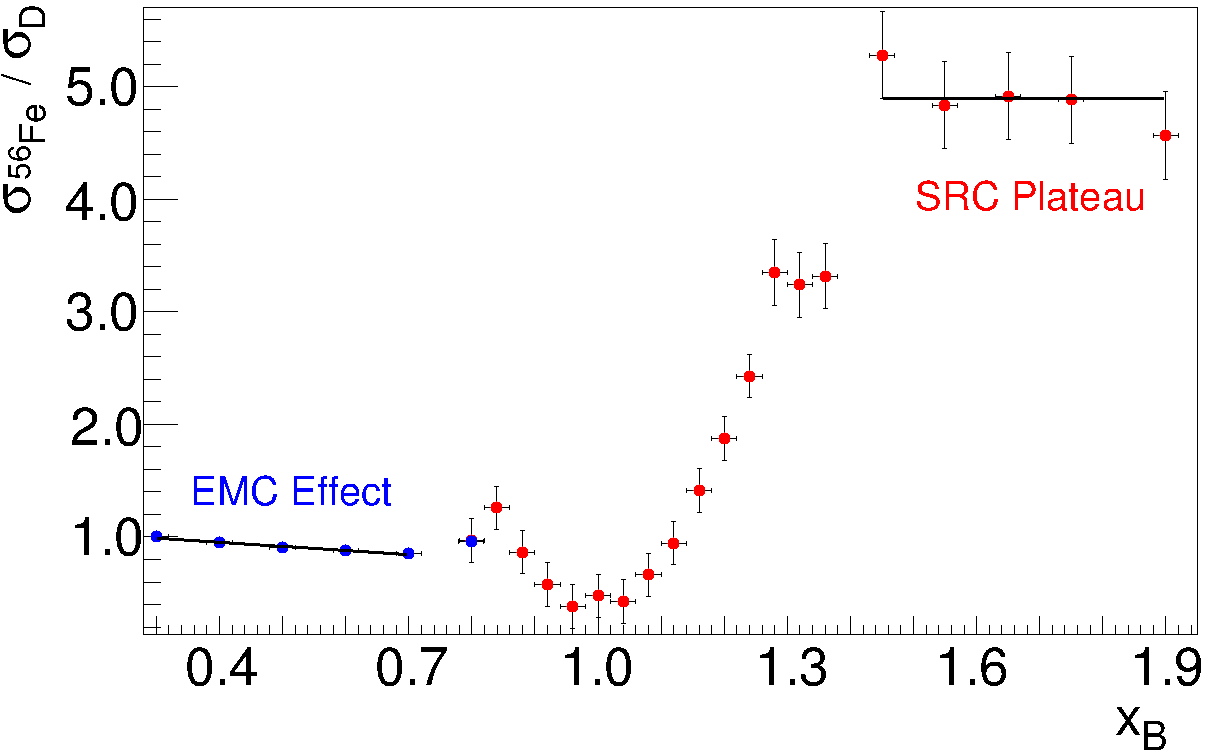}
}
\qquad
\begin{minipage}{6.2cm}
\includegraphics[width=6.2cm]{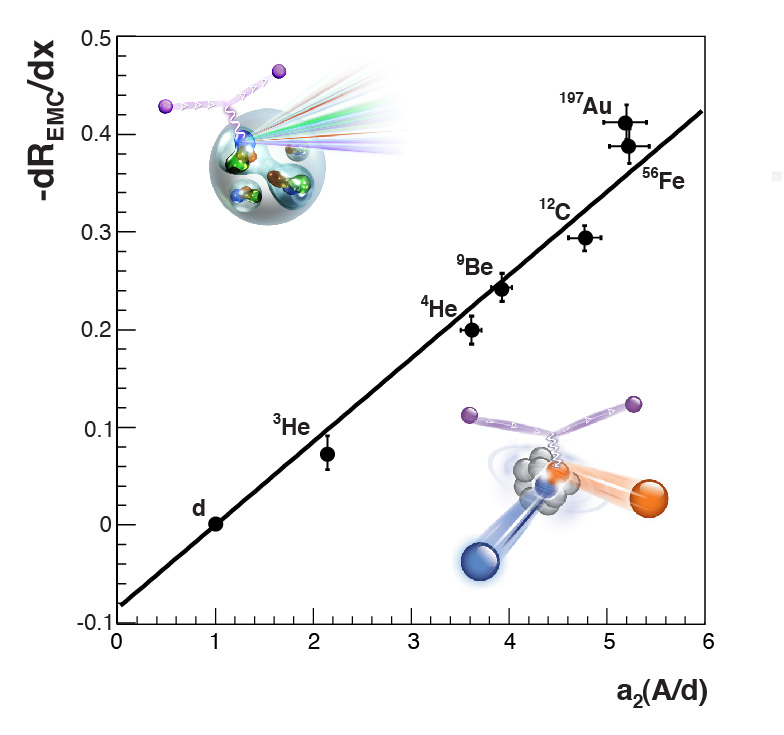}
\end{minipage}
\caption{{\em (left panel)}  The Bjorken x dependence of the ratio of inclusive ($ee^\prime$) cross section of $^{56}$Fe to the deuteron targets. 
Figure adapted from Ref.\cite{Higinbotham:2013hta}.
{\em (right panel)} Observed correlation between the strengths of the EMC effect  and 2N SRCs. Figure used with permission from Jefferson Lab.
Figure adapted from Ref.\cite{Hen:2013oha}.}
\label{fig:emc_src}
\end{figure}

In 2011, L. Weinstein~{\it{et al.}}~\cite{Weinstein:2010rt} noted a linear relation between the magnitude of
the EMC effect and the magnitude of the aforementioned inclusive high-momentum plateaus
as shown in Fig.~\ref{fig:emc_src}.  There is a clear linear relationship between these two seemingly disconnect phenomena which very well
 may be due to the short-range behaviors of the proton-neutron pairs in the nucleus~\cite{Hen:2016kwk}.
 Later publications, which included more data, saw similar
behaviors, though the data alone do not provide a clear cause for the relationship~\cite{Arrington:2012ax,Hen:2012fm}.

A modern review on the topic of nucleon-nucleon corrections and the EMC effect points
to the fact that the strongly interacting proton-neutron
pairs which have been shown to be a universal aspect of high momentum nucleons in the nucleus may
also be responsible for the modification of the structure functions.   For a complete discussion
of the possible connections between nucleon-nucleon correlations and quarks  see
the recent  review by O.~Hen~{\it{et al.}}~\cite{Hen:2016kwk}.

\subsection{Recent (e,e'p) Measurements}

Although  nuclear theory  long ago identified  the need to include high-momentum components to nuclear
momentum distributions and all modern nucleon-nucleon potentials generate  high momentum
tail in the nuclear wave function with a strength far beyond what one would expect from an independent particle models;
it is nevertheless quite challenging to probe directly the high momentum  component of the nuclear wave function.
The above mentioned inclusive results as well as elastic nuclear scattering at large momentum transfers  give a very strong indication 
of initial-state SRCs in the nuclear wave function allowing also to estimate their overall strength. However these processes did not 
allow to extract the shape of high momentum distribution allowing to probe only the integrated characteristics of SRCs.
Thus a large experimental  effort was put in place to measure knock-out (e,e'p)  reactions to determine 
the shapes of high momentum distributions at large  values of residual nuclear excitation energies.

Early nucleon knock-out experiments, with (e,e'p) missing momenta less then the Fermi
momentum were successfully able to extract the momentum distribution of nucleon in the 
nucleus from the measured cross sections~\cite{Ducret:1993wa}. It was also
conjectured at the time, that by simply pushing to larger momentum transfer, 
it would be possible to extract the high momentum part of the momentum distribution; but here nature
would not be so kind.   As shown in many experiments, reaction mechanisms,
final-state interactions or virtual nucleon excitations could quickly dominate the high momentum
signal~\cite{Blomqvist:1998fr} making it extremely challenging to determine the
initial-state high momentum distribution.

Part of the problem was the limited kinematics reach of the accelerators at the time.  
This limited the high missing momentum data to the region between the quasi-elastic peak
and the delta resonance.    A region commonly known as the "dip" region.   With the advent
of JLab, the first $(e,e^\prime p)$ experiments have been measured 
at large missing momentum on the quasi-elastic peak ($x\approx 1$).  The kinematics 
of these measurements corresponded to the large transverse component of  missing momentum,
resulting in a  strong dominance of  final-state interactions  in  the measured cross sections
~\cite{Ulmer:2002jn,Benmokhtar:2004fs,Rvachev:2004yr}.
 
By pushing the kinematics to $Q^2$ = 3.5 GeV/c$^2$ and making use of the previous result to minimize 
final state interaction effects it has been possible to isolate the high momentum component of the deuteron with minimum
model dependence~\cite{Boeglin:2011mt}.
Future (e,e'p) experiments at Jefferson Lab with a 12~GeV beam will exted these kinematics even further, going to
both large missing momentum, $Q^2  >>$ 1, and $x_B > 1$~\cite{Boeglin:2014aca}.   
These kinematics will ideally satisfy the conditions of Eq.(\ref{highkin}) allowing an 
access to the momentum distribution of the deuteron at unprecedentedly large values.
Asymmetry measurements have also been useful in  isolating SRC effects, though to date 
the unambiguous  interpretation of the data require  inclusion of several effects 
associated with long range two-body currents ~\cite{Passchier:2001uc, Mayer:2016zmv,Mihovilovic:2014gdi}.

\subsection{Triple Coincidence Processes}
One of the most important recent advances in studies of the structure of SRCs has been made in 
triple coincidence experiments~\cite{piasetzky06,shneor07,subedi08,Hen:2014nza}.  The possibility of 
reaching high enough momentum transfer that allows to distinguish between struck nucleon from the 
nucleon recoiled from the SRC allows to gain important information about the dynamics and the 
composition of SRCs.

\subsubsection{Angular Correlation of Nucleons in the 2N SRC}
The kinematics of these experiments were close to the  one discussed in Eq.(\ref{highkin}) in which
the detected recoil nucleon with momentum $p_r$ can be associated with the spectator nucleon.

From the theory point of view, such a reaction within PWIA will be described as:
\begin{equation}
{d\sigma\over d\Omega_{e^\prime}dE_{e^\prime} d^3p_N d d^3p_r} =
{F_N\over F_A}\sigma_{eN}\cdot D_A(p_i,p_r,E_r) ~,
\label{eeNNr}
\end{equation}
where the decay function $D_A(p_i,p_r,E_r)$~\cite{frankfurt88,sargsian05, Frankfurt:2008zv},
describes the probability that after a nucleon with momentum $p_i$ is
instantaneously removed from the nucleus, the residual (A-1) nuclear  system
will have residual energy  $E_R  = q_0-T_{f}$ and contain a nucleon in the nuclear decay products, with
momentum $p_r$. Note that if factorization of the nucleon electromagnetic current is justified, then the
decay function can be generalized within the distorted wave impulse approximation~(DWIA), which will 
include effects due to final state interaction of outgoing nucleons with the residual nucleus. 
One advantage of high energy kinematics of Eq.(\ref{highkin}) is the emergence  of the eikonal regime in which case 
FSI effects can be isolated to interfere minimally with the SRC signatures\cite{Sargsian:2004tz,sargsian05}.

If now a  2N SRC is probed in $A(e,e^\prime, N_f,N_r)X$ reaction the prominent signatures will be that the decay
function will exhibit a strong correlation between ${\bf p_i} = {\bf p_f} - {\bf q}$ and ${\bf p_r}$ such that 
\begin{equation}
\vec p_i \approx - \vec p_r ~,
\label{pi_pr}
\end{equation}
if both $p_i$ and $p_f > k_F$. This relation indicates a strong angular correlation between  the direction of the yield of 
recoil nucleons with the direction of the struck nucleon momentum in the initial state.

Such an angular correlation was observed for the first time in the
high momentum transfer  $p+^{12}\mbox{C} \rightarrow p + p + n +
X$ measurement  at  Brookhaven National Lab's E850 experiment~\cite{Aclander:1999fd,
Tang:2002ww}. The experiment measured the  recoil neutrons produced in
coincidence with  high energy proton knockout and observed strong back-to-back  angular correlation 
between direction of the measured recoil neutrons and reconstructed momentum of 
initial bound proton once  these momenta exceeded $k_F$. Remarkably, no correlation was 
observed when reconstructed momentum $p_i< k_F$.

The existence of such  correlations between nucleons in the 2N SRC was confirmed by the  
JLab, Hall A experiment\cite{shneor07} measuring   $^{12}$C(e,e$^\prime$,p$_f$,N$_r$) reaction in  which 
struck proton, $p_f$ has been measured in the coincidence with either recoil proton or neutron, $N_r$.
The kinematics of the experiment  were set at 
$Q^2 \approx 2$~GeV$^2$ and $x \approx 1.2$, with missing momenta in  
300--600~MeV/c.  
With kinematic condition of Eq.(\ref{highkin}) satisfied the experiment observed a clear signature 
for the correlation between the  strength of the cross section and the relative angle 
($\gamma$) of  initial, $\bf p_i$ and recoil nucleon, $\bf p_r$ momenta.

\begin{figure}[htb]
\centering
\parbox{6cm}{
\includegraphics[width=6cm,height=4cm]{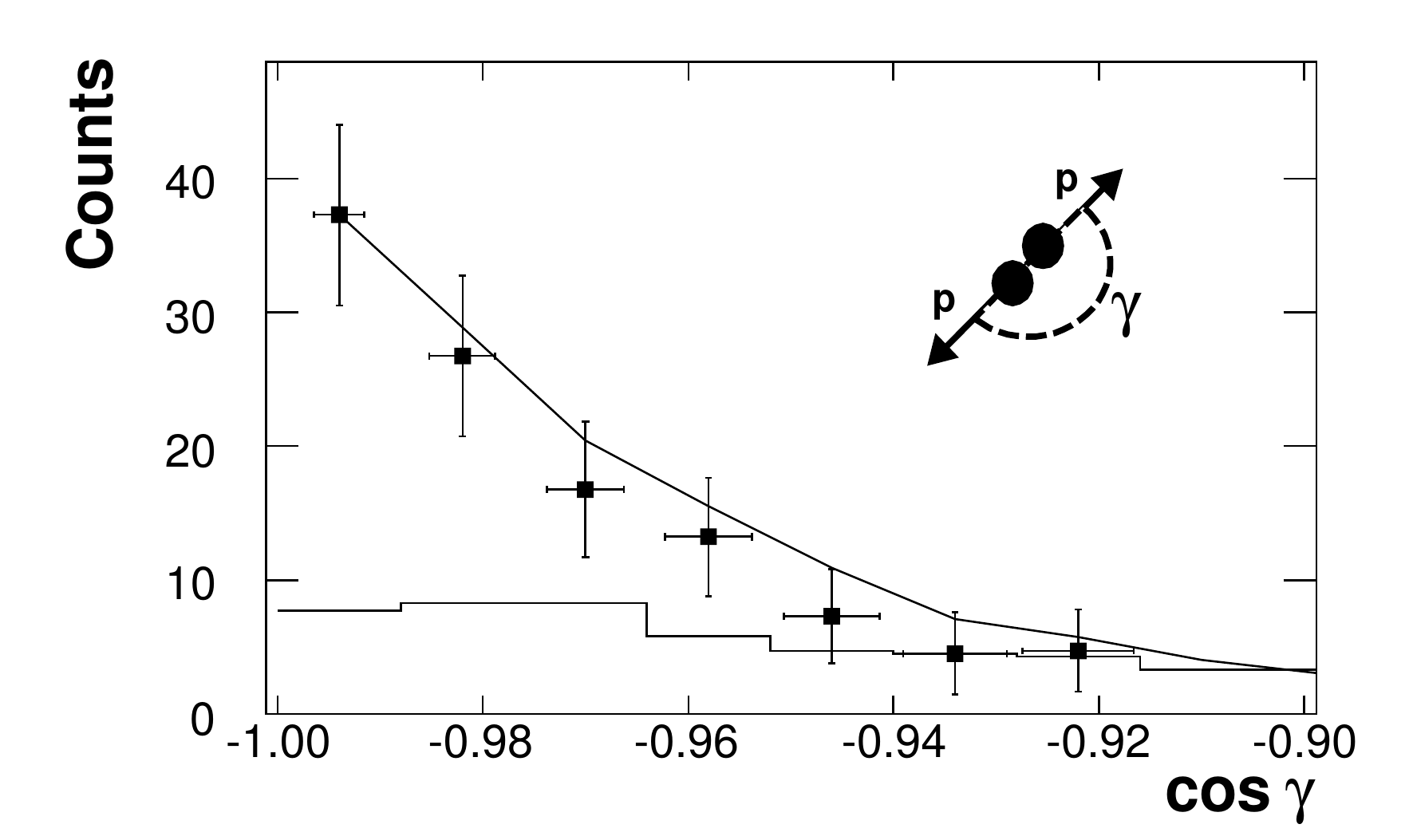}
}
\qquad
\begin{minipage}{6cm}
\includegraphics[width=4cm,angle=270]{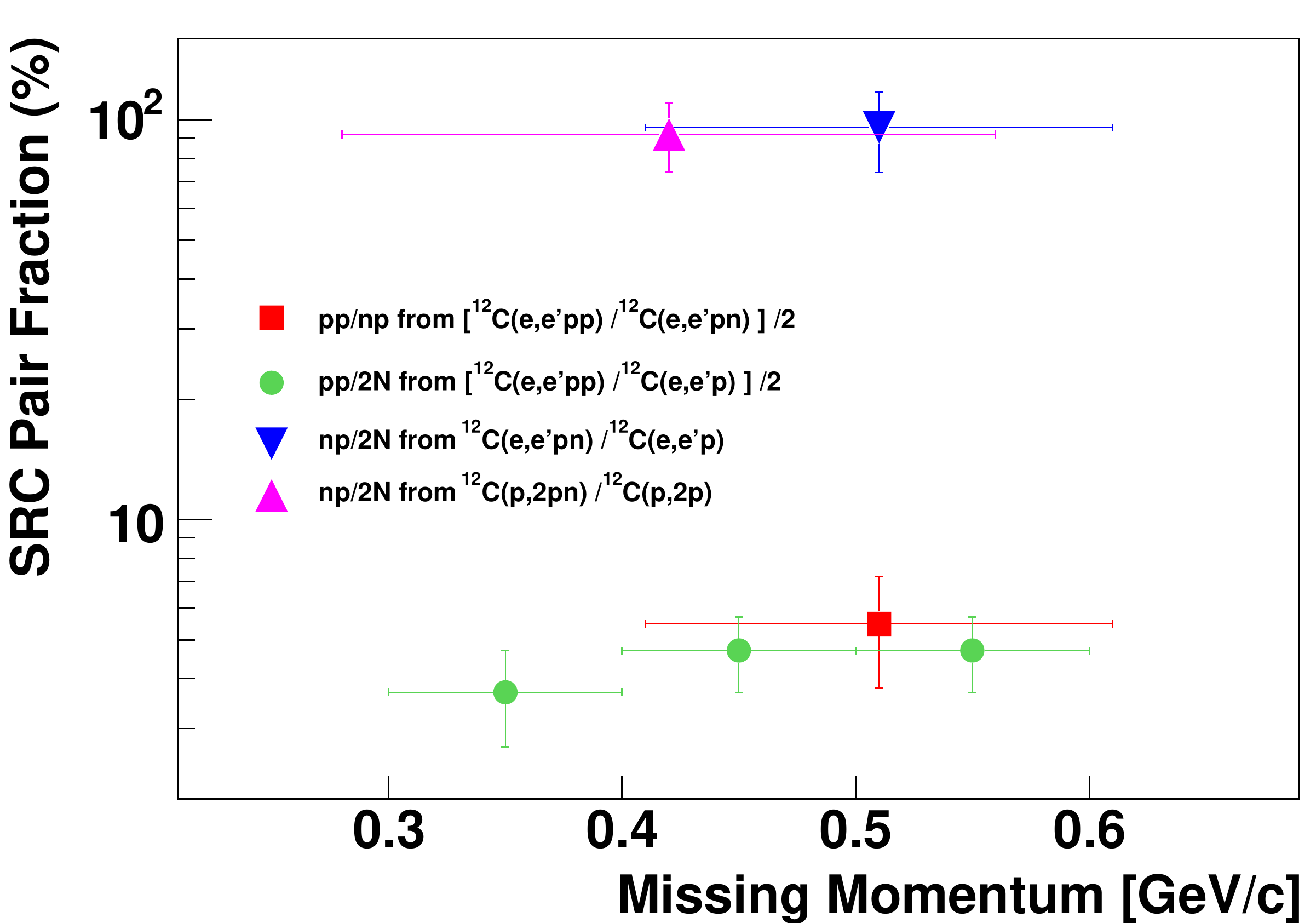}
\end{minipage}
\caption{{\em (left panel)} The distribution of the cosine of the opening angle
between the $\vec p_{i}$ and $\vec p_{r}$ for the $p_i=0.55$~GeV/c for 
the $^{12}$C(e,e$^\prime$pp) reaction. Figure adapted from Ref.\cite{shneor07}. {\em (right panel)} The fraction of correlated pair combinations
in carbon as obtained from the A(e,e$^\prime$pp) to A(e,e$^\prime$pn)
reactions~\cite{subedi08}, as well as from previous $(p, 2pn)$
data~\cite{piasetzky06}.  Figure adapted  from Ref.\cite{subedi08}.}
\label{fig:triple}
\end{figure}

Figure~\ref{fig:triple}(left panel) shows the distribution of
events in $\cos{\gamma}$ for the highest $p_i$ setting of
550~MeV/c~\cite{shneor07}, which is strongly peaked near
$\cos{\gamma}=-1$, corresponding to the back-to-back initial momenta of the
struck and recoil protons.  The solid curve is the simulated distribution for
scattering from a moving pair, with the pair center-of-mass momentum taken to
be a gaussian distribution with a width of 0.136~GeV/c. This width was 
consistent with the one deduced from the $(p, ppn)$ experiment at BNL~\cite{Tang:2002ww} as well
as with the theoretical calculations\cite{CiofidegliAtti:1995qe}. Also shown
in Fig.~\ref{fig:triple}(left panel) is the angular correlation for the random
background as defined by a time window offset from the coincidence peak, which
shows the effect of the acceptance of the spectator proton detector.

\subsection{Observation of pn Dominance in $^{12}C$}
In addition to the observation of strong angular correlation between recoil neutron and struck proton momenta emerging from 
2N SRCs,  the  $p+^{12}\mbox{C} \rightarrow p + p + n + X$ experiment~\cite{Tang:2002ww} determined that 
the  ($49\pm13$)\% of the events with  fast initial protons have  correlated backward going fast recoil neutrons. 

The theoretical analysis of the above experiment~\cite{piasetzky06},  based on the modeling of  the nuclear decay function of 
Eq.\ref{eeNNr}~\cite{Yaron:2002nv},
allowed to relate the above measured event rate  to the quantity $P_{pn/pX}$, which
represents the probability of finding a $pn$ correlation in the
''$pX$'' configuration which contains at least one proton
with $p_i> k_{Fermi}$, yielding:
\begin{equation}
P_{pn/pX} = 0.92^{+0.28}_{-0.18} .
\label{P_pn_exp_BNL}
\end{equation}
This result indicates that the removal of a proton from the nucleus with
initial momentum $275-550$~MeV/c,  in $\sim 92\%$ of the time, is accompanied 
by the  emission of a correlated neutron that carries momentum roughly equal and opposite to the
initial proton momentum.

The BNL experiment did not measure the recoil protons.  However  
the ratio of probabilities of pp and pn SRCs has been  
estimated  in Ref.~\cite{piasetzky06}  using the result of Eq.(\ref{P_pn_exp_BNL}), yielding:
\begin{equation}
P_{pp/pX} \le {1\over 2} (1-P_{pn/pX})= 0.04_{-0.04}^{+0.09}.
\label{P_pp_est_BNL}
\end{equation}

One step forward in verifying the above results on probabilities of $pn$ and $pp$ SRCs was
the above mentioned  Jefferson Lab  experiment\cite{shneor07,subedi08}  in which 
both recoil protons and neutrons  have been detected 
in $^{12}$C(e,e$^\prime$p,N$_r$) reaction.  In this case the experiment  measured the 
fraction of $p_i>k_F$ events in which there was a high-momentum,
backward-angle correlated proton or neutron, i,e,
\begin{equation}
R_{pp} = {N(^{12}C(e,e^\prime p_fp_r)\over N(^{12}C(e,e^\prime p_f)} \ \ \ \ \mbox{and}  \ \ \ \ R_{pn} = {N(^{12}C(e,e^\prime p_fn_r)\over N(^{12}C(e,e^\prime p_f)}
\end{equation}
where $N(\cdots)$ stands for the number of events and $p_f$, $p_r$ and $n_r$ represent struck proton, spectator 
proton and neutron respectively.

 After correcting for the effects  of the detector acceptances and neutron efficiency as well as estimating effects due to 
final state interactions  and absorptions of  produced nucleons the experiment found\cite{shneor07}:
\begin{equation}
R_{pp} = (9.5\pm2)\% \ \ \ \ \mbox{and} \ \ \ \  R_{pn} = 96 \pm 22\%.
\label{R_NNval}
\end{equation}
For $R_{pp}$ it was found to be practically independent of $p_i$ in the 
range of $300< p_i < 600$~MeV/c, while $R_{pn}$ was estimated for 
the  whole  $p_i>300$ range.

In relation to the BNL experiment one observes remarkable agreement between $R_{pn}$ and $P_{pn/pX}$ (of Eq.(\ref{P_pn_exp_BNL})), 
in which both represent the probability of finding a neutron in the $pX$ correlation. 

In the estimation of the probability of $pp$ correlation in the JLab measurement,
the experiment triggered only on forward $^{12}$C(e,e$^\prime$p) events.  Thus  
the probability of detecting pp pairs was twice that of pn pairs, which indicates that for single $pp$  probability one should 
compare $P_{pp} = R_{pp}/2$ to the BNL's estimate of $P_{pp/pX}$,  again observing very good agreement.

The above estimates  result in the following  ratio of the probabilities of   pp to pn
two-nucleons SRCs:
\begin{equation}
{P_{pp}\over P_{pn}} = 0.056^{+0.021}_{-0.012} ~,
\label{pp_pn_JLab}
\end{equation}
which confirms BNL's observation of the strong dominance of $pn$ component in the NN SRCs.

The combined results from BNL and JLab analyses are  presented in Fig.\ref{fig:triple}(right panel), which shows very consistent 
results from BNL and JLab experiments.  The fact that these two experiments employing different probes and covering 
different kinematics in momentum and energy transfer  obtained very similar results  convincingly indicates that 
the observed phenomenon is the genuine property of the nuclear ground state.

\subsection{Observation of $pn$ Dominance in Heavy Nuclei}
The experimental observation of the  $pn$ dominance  in   the $^{12}$C  nucleus   is understood based on the 
dominance of the tensor interaction in the NN SRC.  This fact  itself indicates that the above discussed experiments probed 
NN correlations at  internal separations of $\ltorder 1$~Fm. 
There is rather high confidence in  the validity of this conclusion  
since, as it was discussed in Sec.\ref{sec_pn_dom},  the hypothesis of the $pn$ dominance 
results in two new properties (Eqs.(\ref{p=n}) and (\ref{highn}))   of high momentum component of nuclear wave function which 
are in agreement with the ab-initio variational Monte Carlo calculations of light-nuclei  for up to $^{11}B$.
Thus one will naturally expect that the  $^{12}C$ nucleus being next to $^{11}B$,  will still exhibit the $pn$ dominance. 

If pn dominance is valid also for heavy nuclei it will allow us to extrapolate our results to infinite nuclear 
matter~\cite{McGauley:2011qc} 
with  rather striking implications  for the properties of super-dense asymmetric nuclear matter 
found in the cores of  neutron stars.
However it is not at all obvious that the phenomenon of $pn$ dominance in  NN SRCs will persist for heavy nuclei. 
For example in the  models in which  the high momentum component is generated from correlations between nucleons belonging 
to different nuclear shells  it is predicted that for heavy nuclei,  with an increase of the number 
of  nucleons the contributions from higher orbitals (shells) will increase, increasing the relative strength  of NN SRCs with 
(s=0,T=1) as well as (s=1,T=1) and (s=0,t=0) spin-isospin combinations\cite{Alvioli:2012qa}. 

 Such an  effect can  obscure  the 
dominance of the $s=1$, $t=0$ component  in the short range NN interaction. 
Unfortunately one can not unambiguously verify this question theoretically, since VMC calculations are currently 
applicable for light nuclei only. 
Other ab-initio calculations that address heavy nuclei does not contain short-range interaction component.
 
In this respect the experimental verification of the pn dominance in heavy nuclei is significant.   
Such an analysis of experimental data 
from JLab was performed   in Ref.\cite{Hen:2014nza} covering nuclei of $^{12}$C, $^{27}Al$, $^{56}Fe$ and $^{208}$Pb.
The analysis was similar to the one discussed in the above section~\cite{shneor07}, only that in this case it extracted 
the double ratios for nuclei A relative to 12C, 
 [A(e,e$^\prime$pp)/A(e,e$^\prime$p)] / [$^{12}$C(e,e$^\prime$pp)/$^{12}$C(e,e$^\prime$p)] .  
The data analysis was 
constrained to the  kinematics of $Q^2\ge 1.5$~GeV$^2$ and $x>1.2$  and $300 < p_m < 600$~MeV/c thus 
minimally satisfying condition of Eq.\ref{highkin}.  Similar to Ref.\cite{shneor07} corrections were made to 
account for final state interaction and the absorption effects.
The final result of the analysis demonstrated that, with 
the 65\% confidence level
the $pn$ dominance  is observed for all the nuclei consistent  with the estimate of Eq.(\ref{pp_pn_JLab}).

%\section{New Directions in Short-Range Correlation Studies}
%\input{4new_directions}

\section{New Directions: Probing Short-Range Correlations  in Deep Inelastic Processes}
Theoretically, as first discussed in Sec.\ref{sfquarks}, one can use deep inelastic processes to probe 
super-fast quarks ($x = {A Q^2\over 2 M_A q_0} > 1$) in nuclei.   With the high energy electron beams that
are now available, this is promising new technique for probing nuclear structure at short distances.

\subsection{QCD Evolution of Superfast Quarks}
One way of probing superfast quarks experimentally is 
in deep inelastic scattering from nuclei at $x>1~$\cite{frankfurt88,Sargsian:2002wc}. 
The signature that DIS experiments reached the superfast quark regions comes from extracting nuclear partonic distributions
that satisfy QCD evolution equations.
The first attempt  at JLab to reach the superfast region of nuclear partonic distribution was made with a 6~GeV electron beam~\cite{fomin2010ei}.  
In this experiment, due to the moderate values of  $Q^2\sim 7$~GeV$^2$,  the biggest challenge was to account for  the 
large higher twist as well as finite mass effects.   An interesting aspect of the new measurement was how
it compared with the earlier   measurements from BCDMS/CERN\cite{Benvenuti:1994bb} at $\langle Q^2\rangle \sim 150$~GeV$^2$ 
 and CCFR/FermiLab\cite{Vakili:1999qt} at $\langle Q^2\rangle \sim 125$~GeV$^2$  which yielded  mutually  contradictory results.

The BCDMS collaboration\cite{Benvenuti:1994bb}  measured nuclear structure function,
$F_{2A}(Q^2,x)$ in deep-inelastic scattering of 200~GeV muons from a ${}^{12}$C  target
extracting data for $\langle Q^2 \rangle= 61-150$~GeV$^2$ and $x=0.85-1.15$ ranges.
 The  {\em per nucleon} $F_{2A}$   has been fitted  in the form:
\begin{equation}
F_{2A}(x,Q^2) = F_{2A}(x_0=0.75,Q^2)e^{-s(x-0.75)},
\label{slope}
\end{equation}
with the slope factor estimated as: $s=16.5\pm 0.6$.  Such an exponent corresponds to a very marginal 
strength of the high momentum component of the nuclear wave function.
 
The CCFR collaboration\cite{Vakili:1999qt} extracted {\em per nucleon} $F_{2A}$ for ${}^{56}$Fe  target 
measuring  neutrino and antineutrino scattering in the charged current sector for 
$\langle Q^2 \rangle= 125$~GeV$^2$  and $0.6 \le x \le 1.2$. 
The experiment obtained the slope of the $x$ distribution in the form of Eq.(\ref{slope}), with the exponent 
being evaluated as $s=8.3\pm 0.7 \pm 0.7$. This result was in clear contradiction with the  BCDMS result, requiring 
a much too  large strength of  high-momentum component in the wave function of the ${}^{56}$Fe nucleus.
This strength  was  larger than the one deduced from 
the quasi-elastic electroproduction in the $x>1$ region\cite{frankfurt93,egiyan03,egiyan06,fomin2012,Sargsian:2012sm,McGauley:2011qc}.

The existing contradiction can in principle be solved by the JLab experiment if their $F_{2A}$ could be related to the BCDMS and CCFR data by 
QCD evolution equation.  In the JLab experiment~\cite{fomin2010ei}  
the  {\em per nucleon} structure functions $F_{2A}$  have been  extracted in the $Q^2$ ($6$-$9$~GeV$^2$) range  for 
the ${}^{12}$C target.  Provided these structure function are corrected for finite target mass and higher twist effects 
they can be used as an input to the QCD evolution equation to relate them to the structure functions at different 
$Q^2$ range.  One important feature  of the high $x$ kinematics is that, due to the negligible  contribution
from gluons, the evolution equation for $F_{2A}$ at given $Q^2$  is fully expressed through the input of the same 
structure function measured at different  $Q^2$, i.e.\cite{Freese:2015ebu}:
\begin{eqnarray}
  {d F_{2A}(x,Q^2)\over d \log Q^2}  & = &    {\alpha_s\over 2\pi}\left\{ 2\left(1+{4\over 3}\log\left(1-{x\over A}\right)\right) F_{2,A}(x,Q^2) \right.
  \nonumber \\ 
    & + &   {4\over 3}\int\limits_{x/A}^1{dz\over 1-z}\left({1+z^2\over z}F_{2A}\left({x\over z},Q^2\right) - 2 F_{2A}(x,Q^2)\right) .
  \label{F2A_eveq}
\end{eqnarray}

\begin{figure}[hbt]
\centering
\vspace{0.6cm}
\includegraphics[angle=90,width=12cm,height=3.2cm]{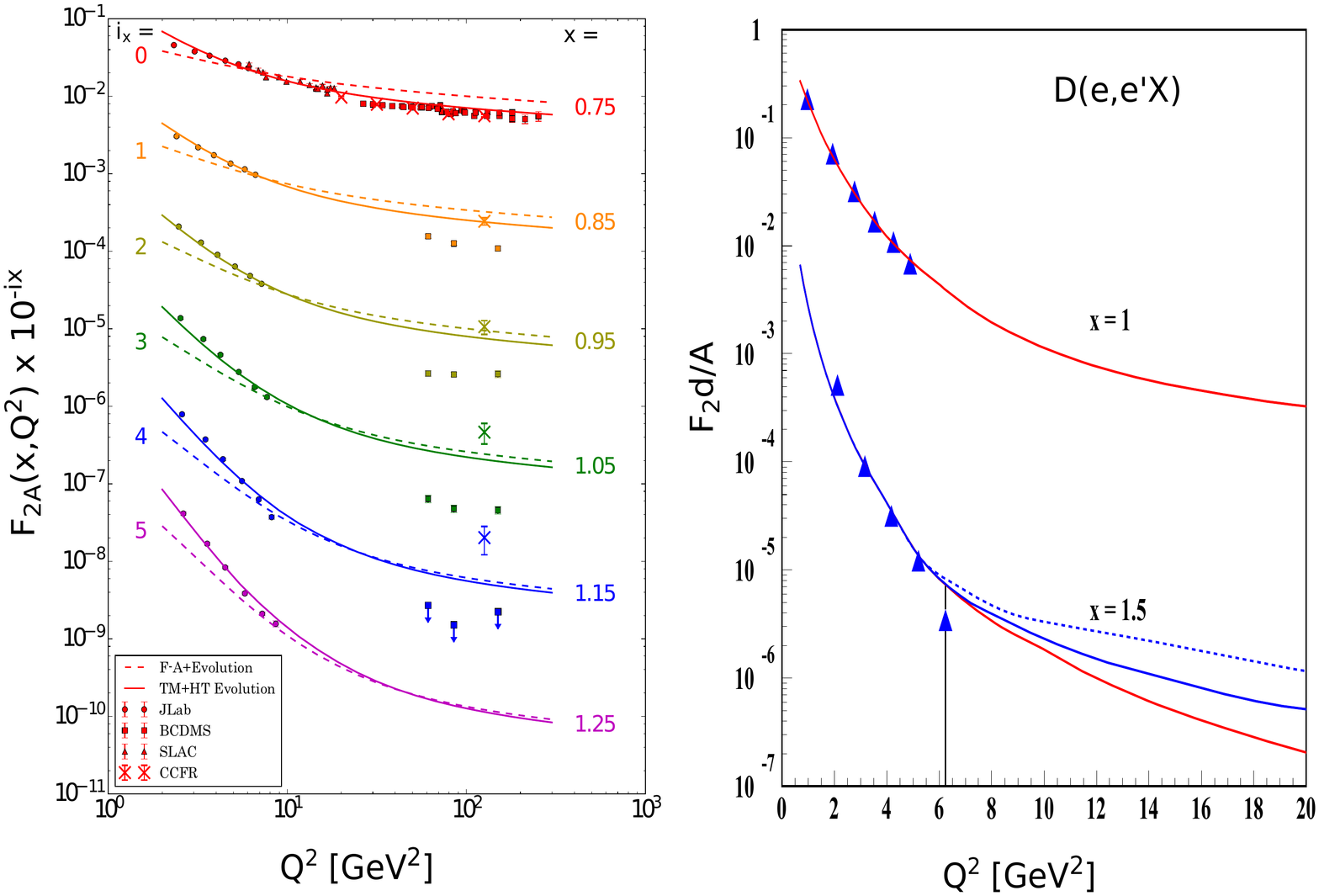}
\vspace{1.2cm}
\caption{{\em (left panel)} Comparison of evolution equation results for the {\em per nucleon} $F_{2A}$ of $^{12}$C 
    to experimental measurements. The JLab data are the ones discussed in the text, while details 
    on CCFR, BCDMS and SLAC data are given in Ref.\cite{fomin2010ei}.
    The structure function is multiplied by $10^{-i_x}$ in order to separate the curves;
    the values of $i_x$ for each $x$ value are given in the plot.  {\em (right panel)} 
    The DIS structure function of the deuteron is calculated with the convolution model with nucleon 
    modification (lower solid line), with the six-quark model (dotted line) and with the hard gluon exchange model 
(upper solid line). (The above descriptions are related to the curves  
labeled by  $=1.5$). Data are from~\protect\cite{Arrington:1998ps}. }
\label{fig:sfquarks}
\end{figure}

In Fig.\ref{fig:sfquarks}(left panel)
the results of the evolution of JLab $F_{2A}$ to the region of $Q^2$ of BCDMS and CCFR experiments are given. 
Two curves correspond to the two different procedures of the extraction of the leading twist part of the 
structure function that is used as an input to the evolution equation of Eq.(\ref{F2A_eveq}). In the one, labelled as 
F-A evolution, the experimental $F_{2A}$ is corrected for the  finite target mass effects according to Ref.\cite{Schienbein:2007gr} 
and parameterized  for fixed  $Q^2=7$~GeV$^2$ (for details see Ref.\cite{fomin2010ei}).  In the second case, labelled as TM+HT evolution,   
the same JLab data have been analyzed simultaneously for finite target mass and higher twist effects.  
The finite target mass was accounted 
for by employing the Nachtmann variable $\xi = 2x/(1+\sqrt{1+Q^2/\nu^2})$ while higher twist effects separated through 
the parameterization of raw data in the form of the inverse powers of $Q^2$ (for details see Ref.\cite{Freese:2015ebu}).

As the comparisons with CCFR and BCDMS data in Fig.\ref{fig:sfquarks}(left panel)  show, the two inputs predict very similar results in
the high $Q^2$ domain, agreeing better with the CCFR (large high momentum nuclear component case)  data for $x\le 1.05$ 
and $Q^2=125$~GeV$^2$.  For $x\sim 1.15$ evolution of JLab structure functions 
predict a  $F_{2A}$ somewhat   in between CCFR and BCDMS data.   This result indicates that the actual strength of the 
high momentum component of nuclear wave function is likely  larger than BCDMS prediction and smaller than that of CCFR.

Again, it is worth emphasizing that the QCD evolution equation provides the best signature that 
the DIS process probed the superfast quarks in nuclei and  gives completely new tool for 
probing the strength of the high momentum component of nuclear wave function.
By using the evolution equation to relate measurements in different $Q^2$ domains at
at $x>1$, the new JLab 12~GeV data will provide a clear indication whether  the deep inelastic 
scattering has probed superfast quark distributions.

\subsection{The Dynamics of the Generation of Superfast Quarks}
As discussed above, the onset of the DIS regime in which 
superfast quarks are probed will result in a unique relation between stricture functions $F_{2A}$ measured at 
different $Q^2$ regions by the QCD evolution equations. 

However, these relations does not allow one to identify the dynamics responsible for the  generation of 
superfast quarks in the nuclear medium.  From the discussion of the kinematics of DIS at $x>1$ in 
Sec.\ref{sfquarks}  one observes that with an increase of $Q^2$ one can reach a  domain of incredibly 
large internal momenta in the nucleus (Fig.\ref{fig:SRCkin}(left panel)).   In this respect, one of the 
main questions is whether the 
nucleonic degrees of freedom are still relevant for the description of the process. 
To address this issue one can consider significantly different models in the generation of the superfast quarks 
and then investigate the feasibility of their experimental verification. 

\subsubsection{Convolution Model} The first most conventional approach is the convolution model.
In convolution model it is assumed that the short 
range correlation of two or more nucleons in the nuclear medium provide 
sufficient initial momentum to the bound nucleon.   These nucleons in turn supply the 
necessary momentum fraction to the  superfast quarks.   Within this scenario nucleons retain
their degrees of freedom but their structure may be strongly modified. 
Thus, in this case the nuclear structure function, $F_{2A}$ is expressed 
through the convolution of structure function of bound nucleon, $F_{2N}$,  and 
nuclear density matrix, $\rho^N_d(\alpha_i,p_t)$,  as follows~\cite{frankfurt88,Sargsian:2002wc,Sargsian:2001gu,Melnitchouk:1996vp}:
\begin{equation}
F_{2A}(x,Q^2) = \sum\limits_{N}\int\limits_{x}^2 \rho^N_A(\alpha_i,p_t)
\tilde F_{2N}({x\over \alpha},Q^2)
{d\alpha\over \alpha} d^2p_t,
\label{convmod}
\end{equation}
where  $\alpha$ represents the light cone 
momentum fraction discussed earlier and $p_t$ is the 
bound nucleon's transverse momentum. The bound nucleon 
structure function, $\tilde F_{2N}$ should account for the 
nuclear medium modifications in agreement with the
EMC effect (see e.g. \cite{frankfurt88,Sargsian:2001gu,Melnitchouk:1996vp}).

This model can be considered rather conventional since except to 
the nuclear EMC effect no non-nucleonic degrees of freedom is invoked  in the calculation. 
In addition to the medium modification effects which one expects to be  proportional 
to the internal momentum of the bound nucleon, the short range phenomena here enters
through the  $\rho^N_d(\alpha_i,p_t)$ function, which contains all the effects of SRCs. The  nuclear 
core effects here enters through the $NN$ potential and one expects that for hard core potentials it 
will result in the fast vanishing $\tilde F_{2A}$ in the $x>1$ region at large $Q^2$.

\subsection{Six-Quark Model}

Six-quark model is an  extreme approach in the description of 
the evolution of superfast quarks in the nuclear medium in the region of $x\ltorder 2$.  
In this case one assumes that short-range interaction   between six quarks 
are responsible for the generation of  superfast quarks.  In the typical 
diagram the six colinear quarks will exchange five  hard gluons  transferring 
the large part of the total momentum fraction to 
the superfast quark which is  subsequently  probed by  the virtual photon.  
  
In asymptotically large $Q^2$ and $x\rightarrow 2$ limit, one can deduce 
the $x$ dependence of the structure function  $F_{2A}$ of nuclei using 
the general quark counting rule~\cite{Brodsky:1973kr} according to which:
\begin{equation}
F_{2A}(x)^{6q}\sim (1-{x\over 2})^{10}.
\label{6q}
\end{equation}
It is important to note that in this model one assumes that the
large momentum fraction carried by the superfast quark is achieved due to 
mixing of all six quarks through the exchanges of hard gluons, thus allowing 
a substantial contribution from the hidden color component of 6q system.
In this case however the two nucleon system is totally collapsed into 6q state 
with complete disappearance of nucleonic degrees of freedom and with no suppression due to 
the phenomenological hard core repulsion. Thus one expects in this case much softer 
$x$ dependence of $F_{2A}$ in the $x>1$ region.  In our numerical 
estimations we use the particular parameterization of 6q model given in 
Ref.\cite{Carlson:1994ga}.

\subsection{Hard Single Gluon Exchange Model}

In this  model we consider rather intermediate scenario in 
which large momentum fraction is supplied to the superfast quark 
not through the mixing of all the six quarks involved in two nucleon 
system, but just by one hard gluon exchange between the two partons, 
belonging to two different nucleons.
In this model only those diagrams contribute  
in which the ``communication'' between two nucleons happens through 
the single hard gluon exchange between two partons.  Such diagrams results in the 
convolution of two partonic distribution functions, one probed by the external photon  and 
the other by the exchanged hard gluon:
\begin{eqnarray}
F_{2A}(x)  & \approx &   N \left[ \int \Psi_{A}(\alpha,p_t) {d\alpha\over \alpha} 
{d^2p_t\over 2(2\pi)^3}\right]^2
\\ \nonumber
& \times & \int\limits_{0}^{1}\int\limits_{0}^{1}(1 - {x\over y_1 + y_2})^2
\Theta(y_1+y_2 - x)f_N(y_1)f_N(y_2) dy_1dy_2,
\label{hgex}
\end{eqnarray}
where $f_N$ is the parton distribution function of the nucleon, and $y_1$ and $y_2$ 
represent the momentum fractions of partons, one from each nucleon, 
participating in the hard scattering.  
In this model some of the effects of short range repulsion will be present in the non-perturbative 
dynamics of parton distribution function of the nucleon. In the asymptotically high $Q^2$ and $x\rightarrow 2$ 
the model has the parametric for of the 6q model due to the $(1-y)^3$ dependence of the nucleon PDFs.

\medskip
In summary we present three different scenarios how the super fast quarks can be generated in the NN system. 
The best way of checking it experimentally is using the deuteron target, since for heavy nuclei multi-nucleon 
SRCs might contribute strongly masking the effects due to the NN interaction at the core.

In Fig.\ref{fig:sfquarks}(right panel) we present the predictions of above discussed models for the 
structure function of the deuteron at $x=1$ and $x=1.5$.
These estimates show the real possibility for discriminating between different scenarios of 
the interaction at core distances in the future experiment with 12 GeV Jefferson Lab.

\medskip
\medskip

Extending the above discussion to heavy nuclei it is worth mentioning that 
in addition to the question how the transition from NN system 
to quark configuration happens one need to  address the question of the 
3N- and higher order SRCs.
An interesting implication of the role of the 3N SRCs  in the DIS regime 
is that  if one considers the ratio of cross sections similar to Fig.\ref{fig:ratios_2n} in $1 < x <2$ region  
the plateau will disappear with the increase of $Q^2$. The observation of 

such an effect is due to the fact that with an increase of $Q^2$  in the DIS regime 
the internal momenta steeply increase at fixed $x$ (Eq.(\ref{pin})) as a result even for $x<2$ 
one should expect substantial contribution due to 3N SRCs which will 
disrupt the plateau observed experimentally in the quasi-elastic kinematics.

%\section{Conclusion and Outlook}
%\input{5conclusions}
\section{Conclusion and Outlook}

We have reviewed  the  recent progress in studies of short range correlations in nuclei
that has been driven
by a the series of high energy experiments with proton and electron probes. 
We demonstrated how the inclusive electronuclear processes in the $x > 1$ quasi-elastic region 
were able to identify NN short range correlations in nuclei and extract the parameter $a_2(A,Z)$ that 
characterizes the strength of NN SRCs in the high momentum part of the nuclear wave functions.
We discuss the observation of apparent correlation between the strength of the medium modification of 
partonic distributions in nuclei and the $a_{2}(A,z)$ factor of 2N SRCs.

For the  triple-coincident experiments, we  reviewed the observed strong angular correlation between 
the constituents of 2N SRCs and the  strong dominance of the $pn$ component in these correlations.
The $pn$ dominance is understood based on the large  tensor interaction in the NN SRCs at $\ltorder 1$~fm 
distances.  We reviewed the implication of this dominance on the properties of the high momentum distribution
of nucleons in 
nuclei and recent observations which apparently are in agreement with these properties.

The next subject of the review was the physics of the three-nucleon correlations where the current results
are rather inconclusive.  We show that in order to make a definitive probe of the three-nucleon correlations
one needs  considerably higher $Q^2$ which can be reached in upcoming experiments at Jefferson Lab.

Finally, we review the new and promising direction of studying  short range properties of nuclei at core distances
with deep inelastic scattering at Bjorken $x>1$ region. 
We reviewed the first such experiment completed at Jefferson Lab and demonstrated its potential 
in verifying the validity of  QCD evolution equation for such superfast quarks.
We discussed also the sensitivity of the cross section of inclusive scattering from the superfast quarks to 
the particular mechanism of $NN$ interaction at core distances.

Future  experiments at JLab, as well as at high energy labs such as LHC, JPARC and possibly EIC, might provide 
significant new information about the dynamics of  NN interactions at the core.

\section*{ACKNOWLEDGMENTS}

$^*$ \ 
During the writing of this manuscript, our good friend and co-author Patricia Solvignon
passed away.    Though very young, Patricia had already had made quite an impact in
the field of nuclear physics and had several approved experiments at Jefferson Lab to
measure the effects of short-range correlations in nuclei.    In fact, Patricia was a
spokesperson on several of the past and future Jefferson Lab experiments described herein.   
Patricia was a good friend and an outstanding scientist and will be sorely missed. 

\ 

This work is supported by the U.S. Department of Energy, Office of Science, Office of Nuclear
Physics grants under contracts DE-SC0013615, DE-AC05-06OR23177, and
DE-FG02-01ER-41172. We are thankful to  our colleagues for collaboration and assistance
in performing the above described research in general and preparation of the current article in particular.

%\bibliography{src_arnps.bib}

\end{document}